\documentclass[twocolumn]{aastex631}

\usepackage[utf8]{inputenc}
\usepackage{xcolor}

\newcommand{\srca}{J1620$-$4927}
\newcommand{\srcb}{J1646$-$4451}
\newcommand{\srcc}{J1832$-$0808}
\newcommand{\srcd}{J1837$-$0616}

\begin{document}

\title{Discovery of Two Highly Scattered Pulsars from Image-Based Circular Polarization Searches with the Australian SKA Pathfinder}

\author[0000-0002-9409-3214]{Rahul Sengar}
\affiliation{Max Planck Institute for Gravitational Physics (Albert Einstein Institute), D-30167 Hannover, Germany \\}
\affiliation{Leibniz Universität Hannover, D-30167 Hannover, Germany\\}
\affiliation{Center for Gravitation, Cosmology, and Astrophysics, Department of Physics, University of Wisconsin-Milwaukee, P.O. Box 413, Milwaukee, WI 53201, USA.\\}

\author[0000-0001-6295-2881]{David L. Kaplan}
\affiliation{Center for Gravitation, Cosmology, and Astrophysics, Department of Physics, University of Wisconsin-Milwaukee, P.O. Box 413, Milwaukee, WI 53201, USA.\\}

\author[0000-0002-9994-1593]{Emil Lenc}
\affiliation{CSIRO Space and Astronomy, PO Box 76, Epping, NSW 1710, Australia\\}

\author[0000-0002-8935-9882]{Akash Anumarlapudi}
\affiliation{Department of Physics and Astronomy, University of North Carolina at Chapel Hill, 120 E. Cameron Ave, Chapel Hill, NC, 27599, USA\\}
\affiliation{Center for Gravitation, Cosmology, and Astrophysics, Department of Physics, University of Wisconsin-Milwaukee, P.O. Box 413, Milwaukee, WI 53201, USA.\\}

\author[0000-0002-5119-4808]{Natasha Hurley-Walker}
\affiliation{International Centre for Radio Astronomy Research, Curtin University, Kent St, Bentley WA 6102, Australia\\}

\author[0000-0002-2066-9823]{Ziteng Wang}
\affiliation{International Centre for Radio Astronomy Research, Curtin University, Kent St, Bentley WA 6102, Australia\\}

\author[0000-0002-4405-3273]{Laura Driessen}
\affiliation{Sydney Institute for Astronomy, School of Physics, The University of Sydney, New South Wales 2006, Australia}
\affiliation{ARC Centre of Excellence for Gravitational Wave Discovery (OzGrav), Hawthorn, VIC 3122, Australia}

\author[0000-0003-0699-7019]{Dougal Dobie}
\affiliation{Sydney Institute for Astronomy, School of Physics, The University of Sydney, New South Wales 2006, Australia}
\affiliation{ARC Centre of Excellence for Gravitational Wave Discovery (OzGrav), Hawthorn, VIC 3122, Australia}

\author[0000-0002-2686-438X]{Tara Murphy}
\affiliation{Sydney Institute for Astronomy, School of Physics, The University of Sydney, New South Wales 2006, Australia}
\affiliation{ARC Centre of Excellence for Gravitational Wave Discovery (OzGrav), Hawthorn, VIC 3122, Australia}


\

\begin{abstract}
We report the discovery and timing of two pulsars from a sample of four circularly polarized sources identified in radio continuum images taken as part of the Australian SKA Pathfinder (ASKAP) Variables and Slow Transients (VAST) survey. Observations with the Parkes (Murriyang) radio telescope confirmed both sources as normal pulsars with high dispersion measures. PSR J1646$-$4451 has a spin period of 217\,ms and a dispersion measure (DM) of 928\,$\rm cm^{-3} \, pc$, while PSR J1837$-$0616 exhibits a spin period of 118 ms and a DM of 793\,$\rm cm^{-3} \, pc$. These pulsars show extreme pulse broadening due to scattering, with measured scattering timescales of 117\,ms and 75\,ms at observing frequencies of $\sim$ 1.8 GHz, respectively. These measurements imply extrapolated scattering timescales at 1\,GHz of $\sim$1346\,ms and 740\,ms, placing them among the most heavily scattered known pulsars. Our findings underscore the potential of using circular polarization in radio continuum images as a tool for identifying highly scattered pulsars. Future wide-field radio continuum surveys are poised to uncover a broader population of extreme pulsars, particularly those that are heavily scattered at 1.4\,GHz, intrinsically faint, or residing in binaries--offering valuable insights into both pulsar demographics and the complex structure of the ionized interstellar medium.

\end{abstract}


\section{Introduction} \label{sec:intro}

The majority of pulsars have been discovered through time-domain-based, untargeted, and large-scale pulsar surveys that systematically scan wide regions of the sky in an unbiased manner \citep[e.g.,][]{manchester_01, cordes_06, keith_10, keane_18, han_21, padmanabh_23, stovall_14, sanidas_19}. The dedispersed time-series data are then searched for periodic signals using Fast Fourier Transform \citep[e.g.,][]{handbook_04} or Fast Folding Algorithm \citep{staelin_69} techniques. However, these searching algorithms result in hundreds of millions of candidates, the majority of which arise from radio frequency interference (RFI), noise, or statistical fluctuations that can mimic genuine pulsar signals. This necessitates substantial post-processing and candidate vetting efforts. If, however, one could \textit{a priori} identify sources in the sky that are likely to be promising pulsar candidates, and subsequently conduct time-domain searches specifically for those sources, it could potentially yield significant gains in both computational efficiency and detection sensitivity. Such a targeted refinement of the standard blind-search approach could result in interesting pulsars while simultaneously reducing the volume of spurious candidates.

Radio continuum imaging surveys provide a complementary approach to pulsar searches by targeting their persistent, steep-spectrum emission \citep{jankowski_18}, rather than relying solely on time-domain periodicity. This approach is particularly advantageous in scenarios where traditional time-domain searches are limited by propagation effects in the interstellar medium (ISM). When pulsar signals pass through an inhomogeneous interstellar medium, they suffer from temporal smearing due to multipath scattering \citep{sutton_ism_71}. Unless the pulsar is sufficiently bright, its signal may be lost in background noise due to scattering. Observing pulsars at higher frequencies can mitigate this effect as the scattering timescale decreases with increasing frequency, i.e., $\tau \propto \nu^{-4}$ \citep{rickett_77}. However, due to the steep spectrum nature of the majority of pulsars and lower frequencies providing a larger field of view and thus faster survey speed, most large-scale time-domain surveys have been conducted at or below 1.4\,GHz \citep[e.g.,][and references therein]{sengar_htru_25}. However, this makes them particularly susceptible to scattering in the Galactic plane. Similarly, orbital modulation in binary pulsars smear out the pulsed signal and can significantly hinder detection if the effects of orbital acceleration are not properly accounted for \citep{johnston_91}. Finally, ephemeral pulsars such as rotating radio transients (RRATs) \citep{maura_06, keane_11}, eclipsing pulsars \citep{thompson_94}, and nulling pulsars \citep{rickett_90, wang_07} also often evade detection due to their sporadic emission. Unlike time-series analyses, continuum observations integrate emission over long durations, effectively mitigating these effects.

\begin{deluxetable*}{lcccccccccc}
\tablecaption{Observed properties of selected sources in VAST. The table lists the coordinates of each source, mean of total intensity ($I_{\mathrm{mean}}$), circular polarization ($V_{\mathrm{mean}}$), fractional circular polarization ($|V|/I$), signal-to-noise ratio (S/N), the $K_s$-band magnitude (AB), and the source type.}
\label{tab:sources}
\tablehead{
\colhead{Source Name} & \colhead{RA} & \colhead{DEC} & 
\colhead{GL} & \colhead{GB} & 
\colhead{$I_{\mathrm{mean}}$} & \colhead{$V_{\mathrm{mean}}$} 
& \colhead{$\rm |V|/I$} & \colhead{S/N} & \colhead{$K_s$} & \colhead{Type} \\
\colhead{} & \colhead{(J2000)} & \colhead{(J2000)} & \colhead{(\degr)} & \colhead{(\degr)} & \colhead{(mJy)} & \colhead{(mJy)} & \colhead{} & \colhead{} & \colhead{(AB)} & \colhead{}
}
\startdata
ASKAP J162048.8$-$492745 & 16:20:48.88(1) & $-$49:27:45(3) & 333.900 &  0.391 &  2.5 &  0.51  & 0.27 &  4.1 & 16.99$\pm$0.01 & star \\
ASKAP J164634.5$-$445126 & 16:46:34.52(1) & $-$44:51:26(3) & 340.250 &  0.305 &  2.4 & $-0.45$& 0.19 &  5.7 & $>$22.2          & pulsar \\
ASKAP J183233.4$-$080835 & 18:32:33.40(1) & $-$08:08:35(3) &  23.538 &  0.453 &  2.3 &  0.47  & 0.26 &  3.3 & $>$22.3          & unknown \\
ASKAP J183720.9$-$061609 & 18:37:20.90(1) & $-$06:16:09(3) &  25.747 &  0.264 & 13.2 &  0.96  & 0.07 & 10.0 & $>$20.8          & pulsar \\
\enddata
\end{deluxetable*}

Radio continuum observations are particularly effective because pulsars are highly polarized sources and exhibit distinctive polarization properties --- typically a high degree of linear polarization (10--50\%) and noticeable circular polarization (5--30\%) \citep[e.g.,][]{wang_23, oswald_23}. These characteristics are rare among other Galactic or extragalactic radio sources such as AGN and radio galaxies, which are dominated by incoherent synchrotron emission with minimal circular polarization \citep[e.g.,][]{saikia_88, Bjornsson_90}. Radio stars, however, show high fractional circular polarization \citep[e.g.,][]{pritchard_21}, but the emission from them is narrowband, transient, and  often linked to stellar counterparts detectable in optical/infrared surveys. Therefore, they can be easily distinguished from the broadband emission of pulsars. The combination of these unique polarization signatures with compact morphology and steep spectral indices at $\sim$GHz frequencies \citep[$\alpha \sim -1.6$, where $S_{\nu} \propto \nu^{\alpha}$;][]{jankowski_18} enables efficient identification of pulsar candidates in wide-field radio continuum images. While such identification does not guarantee a pulsar detection (as other source types may share some characteristics, especially when looking at outliers from large background populations), it significantly narrows down potential targets for follow-up observations. These candidate sources must ultimately be observed with high resolution radio data and analyzed using standard pulsar search techniques (e.g., FFT or FFA) to confirm their pulsar nature. Thus, radio continuum observations serve as a powerful filtering mechanism, converting traditional unbiased searches into more targeted campaigns. This approach is particularly valuable for discovering pulsars in challenging regimes--such as those in highly scattered regions or binary systems--that might otherwise be missed by conventional time-domain surveys. By pre-selecting the most promising circularly polarized candidates, radio continuum imaging can significantly improve the efficiency of pulsar searches.

In the last few decades, the use of radio continuum imaging for pulsar discovery has evolved significantly. The pioneering detection of the first millisecond pulsar PSR~J1939$+$2134 through its characteristic steep-spectrum emission \citep{backer_82} established the potential of this approach long before modern survey capabilities emerged. While early attempts often failed to yield new pulsars, recent advances in radio astronomy instrumentation and interferometers have enabled a new generation of high-sensitivity, wide-bandwidth continuum surveys with unprecedented sky coverage and angular resolution. Facilities like the Australian SKA Pathfinder (ASKAP) with its Evolutionary Map of the Universe (EMU) survey \citep{norris_emu_2011, hopkins_25}, Murchison Widefield Array \citep[][]{lenc_18}, the Karl G. Jansky Very Large Array (VLA) conducting the VLA Sky Survey \citep[VLASS;][]{vlass_20}, Low-Frequency Array (LOFAR) Two-metre Sky Survey \citep[LoTSS;][]{shimwell_17, best_23}, LOFAR's TULIPP project \citep{sobey_22}, and MeerKAT's various survey programs \cite[e.g.,][]{fender_16,padmanabh_23} are producing deep, wide-field radio images ideal for identifying pulsar candidates through their spectral and polarization properties.

Several recent serendipitous discoveries of pulsars illustrate the growing effectiveness of this approach. For example, the identification of the highly polarized ASKAP source as millisecond pulsar PSR J1431$-$6328 previously missed in the High Time Resolution Universe survey due to scattering at L-band frequencies or processing errors \citep{kaplan_19}. Similarly, MeerKAT observations during the 2020 Saturn-Jupiter conjunction serendipitously discovered the strongly scintillating PSR J2009$-$2026 \citep{smirnov_24}. Other examples include the discovery of a pulsar J0523$-$7125 in the Large Magellanic Cloud \citep{wang_22},  discovery of a young pulsar PSR J1032$-$5804 through its circular polarization, despite its high dispersion measure (800 pc cm$^{-3}$) and rotation measure \citep{wang_24}, candidate
redback pulsar binary 4FGL J1646.5$-$4406 \citep{zic_24}, and a highly scattered PSR J1631$-$4722 associated with supernova remnant G336.7$+$0.5 and identified as a compact source (RACS J163159.8--472157) with tail-like emission in the Rapid ASKAP Continuum Survey \citep{mcconnell_20}, with pulsations discovered by  \cite{ahmad_25}. Most recently, a new RRAT \citep{mcsweeney_25} and five eclipsing MSPs \citep{2025PASA...42..139P,2025arXiv251209339P} have been discovered using image-based searches.

These discoveries highlight three critical advantages of continuum imaging: (1) immunity to scattering effects that can limit time-domain searches, (2) sensitivity to polarization properties that serve as unique pulsar fingerprints, and (3) ability to identify pulsars in complex environments like supernova remnants. Statistical analyses now suggest that image-based methods may uncover pulsar populations complementary to those found via periodicity searches \citep{frail_24}. The growing sample of these highly scattered and young pulsars demonstrates that image-based searches will be essential for a complete census of the Galactic pulsar population. Future surveys with the SKA and its precursors are expected to increase the yield of image-discovered pulsars by an order of magnitude, especially in the challenging Galactic plane region where traditional searches are least effective \citep{dai_18}.

In this paper, we present the discovery of two highly scattered pulsars --- PSR J1646$-$4451 and PSR J1837$-$0616 --- which are among a sample of four circularly polarized sources detected in the ASKAP Variable and Slow Transients (VAST) survey \citep{murphy_13,murphy_21} and were subsequently confirmed as pulsars with the Parkes radio telescope. In Section \ref{sec:sources}, we summarize the discovery observations both in the radio continuum data as well as in Parkes radio data. Timing analysis, polarimetry, and scattering analysis are discussed in Section \ref{sec:timing}. The discussion is given in Section \ref{sec:discussion}, and we summarize our conclusions in Section \ref{sec:conclusion}.

\section{Source observations } \label{sec:sources}

\subsection{VAST observations }
\label{subsec:vast_sources}

Since November 2022, the VAST project has been conducting a radio survey using ASKAP which targets the southern Galactic plane. This campaign systematically monitored 41 preselected sky regions, each constrained to Galactic latitudes within $\pm6^\circ$ and declinations below $-10^\circ$ (although the exact selection has changed slightly over time), cumulatively spanning around 1260 square degrees. Observations of these regions were repeated approximately every fortnight. Each session involved a 12-minute exposure at a central frequency of 888\,MHz and employed a bandwidth of 288\,MHz, enabling a median sensitivity of roughly 0.2\,mJy\,beam$^{-1}$ in areas along the plane.

The polarization data used are standard ASKAP data products, in which all instrumental correlations (XX, XY, YX, and YY) are recorded for reconstruction of images in all four Stokes parameters: $I$, $Q$, $U$, and $V$. Data handling was performed offline through the untargeted VAST pipeline, yielding both total intensity and circular polarization images, along with corresponding source catalogs. The bandpass and flux calibration relied on PKS B1934$-$638, and temporal phase variations were corrected via self-calibration procedures. The details of these processing steps are outlined in \citet{murphy_21}.

We focused our analysis on detecting compact sources with high fractional circular polarization, aiming to isolate targets of astrophysical interest, potentially pulsars. After rejecting sources with known multi-wavelength matches, most of which are radio stars \citep{pritchard_21, driessen_24} or known pulsars, we  identified four circularly polarized sources for follow-up (see Table \ref{tab:sources}) whose fractional circular polarization ranges from 7\% to 30\%, similar to what is seen in pulsars \citep{anumarlapudi_23}.

\subsection{Multi-wavelength search for counterparts}\label{sec:mw}

Sources at radio wavelengths that show significant polarization ($\gtrsim5$--10\%) often include stars \citep{pritchard_21}, pulsars \citep{anumarlapudi_23}, and white dwarf binaries, especially magnetic cataclysmic variables \citep[MCVs;][]{barrett2020}. Hence, we searched the archival optical and near--infrared (NIR) data for these sources that might inform us of their nature. At optical wavelengths, we searched the data from \textit{Gaia} \citep{gaiadr3}, the Dark Energy Camera Plane Survey \citep[DECaPS;][where available]{decaps}, and the Panoramic Survey Telescope and Rapid Response System \citep[Pan-STARRS;][where DECaPS data were not available]{panstarrs}. At NIR wavelengths, we searched deep observations from the VISTA Variables in the V\'{i}a L\'{a}ctea Survey \citep[VVV;][]{vvv} and the Galactic Plane Survey from the UKIRT Infrared Deep Sky Survey \citep[UKIDSS;][]{UKIDSS}. Because multiple short exposures are often available in the NIR, we stacked these observations (or used the stacked image, if available) to obtain deeper observations. At the location of these sources, we performed aperture photometry on the stacked images to get source fluxes/upper limits. Zero-point calibration for the stacked images was performed using the DECaPS and UKIDSS source catalogs.

For \srca, we identified the star Gaia DR3 5935425106911225216, $0.6\arcsec$ away from the radio position, well within the ASKAP position uncertainty.  This star has $G$-band magnitude of 18.6, an effective temperature of 4200\,K, and log surface gravity of 4.9. Considering the source density of sources in Gaia DR3 (within a 5\arcmin\ radius), we estimate the probability of random association with \srca\ to be 0.02. \srca\ is also well detected in the DECaPS and VVV data, with the broadband spectral energy distribution (from these colors) consistent with the temperature estimated by the \textit{Gaia} survey. These parameters are similar to many of the K-type stars in the Sydney Radio Star Catalogue \citep{driessen_24}; however, no detectable parallax was measured for \srca. The measured circular polarization ($\sim 27\%$) of this source is high for standard gyrosynchrotron emission from a late-type dwarf, for which one typically expects $\lesssim 20\%$, but it lies within the range observed from magnetically active K-type stars with coherent radio emission, where $|V|/I \sim 30$–$100\%$ has been reported \citep[e.g.][]{pritchard_21,driessen_24}.

For \srcb, no optical counterpart was found, but we found a NIR source in the VVV images $0\farcs83$ away from the radio position. However, considering the higher source density at NIR wavelengths in the Galactic plane, we derive a probability of random association to be 0.4; hence, this association is likely not  robust. Excluding this source, at the position of \srcb, in a 0.5\arcsec\ circle, we estimate an upper limit of $K_s$=22.2 AB.

\begin{figure}
    \includegraphics[width=0.99\columnwidth]{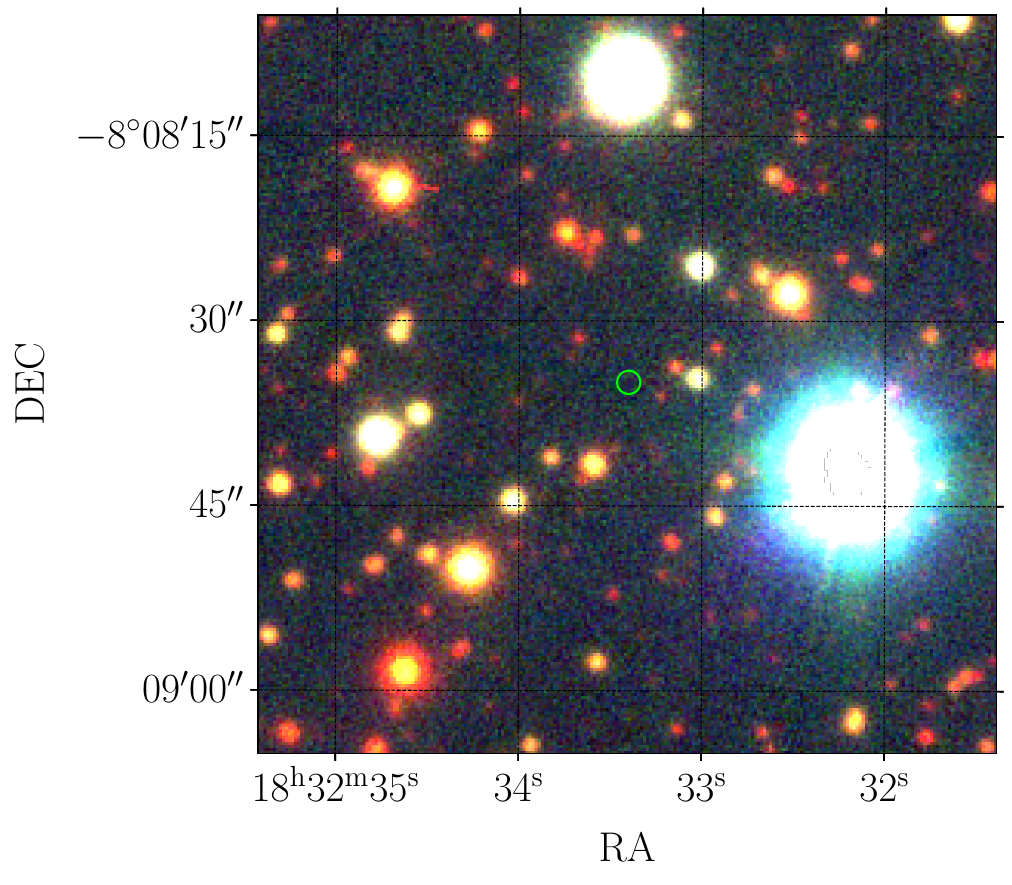}
    
    \caption{Optical \textit{gri} composite image of the field of \srcc, using the data from DECaPS. The image is 30\arcsec\ in extent along both axes, and the lime circle shows the radio position of \srcc.}
    \label{fig:J1832_decaps_image}
\end{figure}

For \srcc, and \srcd, no DECaPS/VVV data were available, and hence we inspected Pan-STARRS and UKIDSS data. In the case of \srcc, no counterpart was found either at optical or NIR wavelengths up to a $K_s$-band limiting magnitude of 22.3 AB (see Figure \ref{fig:J1832_decaps_image}). However, for \srcd, we found a NIR (but not optical) detection of a source, 1.3\arcsec\ away from the radio position. However, similar to \srcb, given the higher source density in the Galactic plane, we estimate the probability of random association to be 0.6 between the NIR source and \srcd. At the location of \srcd, we estimate the limiting magnitude to be $K_s>20.8$ AB.

\subsection{Parkes/Murriyang confirmation observations}\label{subsec:parkes_observations}

In order to confirm whether any of these circularly polarized sources are pulsars, we conducted follow-up observations at the Parkes radio telescope using the Ultra-Wideband Low \citep[UWL;][]{hobbs_20} receiver, which provides continuous frequency coverage from 704~MHz to 4032~MHz. The extensive bandwidth of the UWL system offers enhanced sensitivity to pulsars that may be undetectable at conventional L-band frequencies (e.g., 1000--1500~MHz), either due to low signal significance or strong interstellar scattering. Each source was observed in search mode (total intensity) for durations between 60 and 72 minutes, using a time resolution of 64~\textmu s, 2-bit sampling, and 500~kHz channel bandwidth, yielding a total of 6656 frequency channels. Periodicity searches were performed using the GPU-accelerated \texttt{PEASOUP}\footnote{\url{https://github.com/ewanbarr/peasoup}}
software, which applies time-domain resampling techniques to improve sensitivity to binary pulsars experiencing orbital acceleration. A dispersion measure (DM) range of up to 1500~cm$^{-3}$~pc was searched for each target, corresponding to 1.2--2.0 times the maximum DM predicted by either the NE2001 or YMW16 electron density models. The acceleration search covered a range of $\pm25$~m\,s$^{-2}$, sufficient to detect signals from highly accelerated binary systems. 

\begin{figure*}[t]
    \centering
    \includegraphics[width=0.95\textwidth]{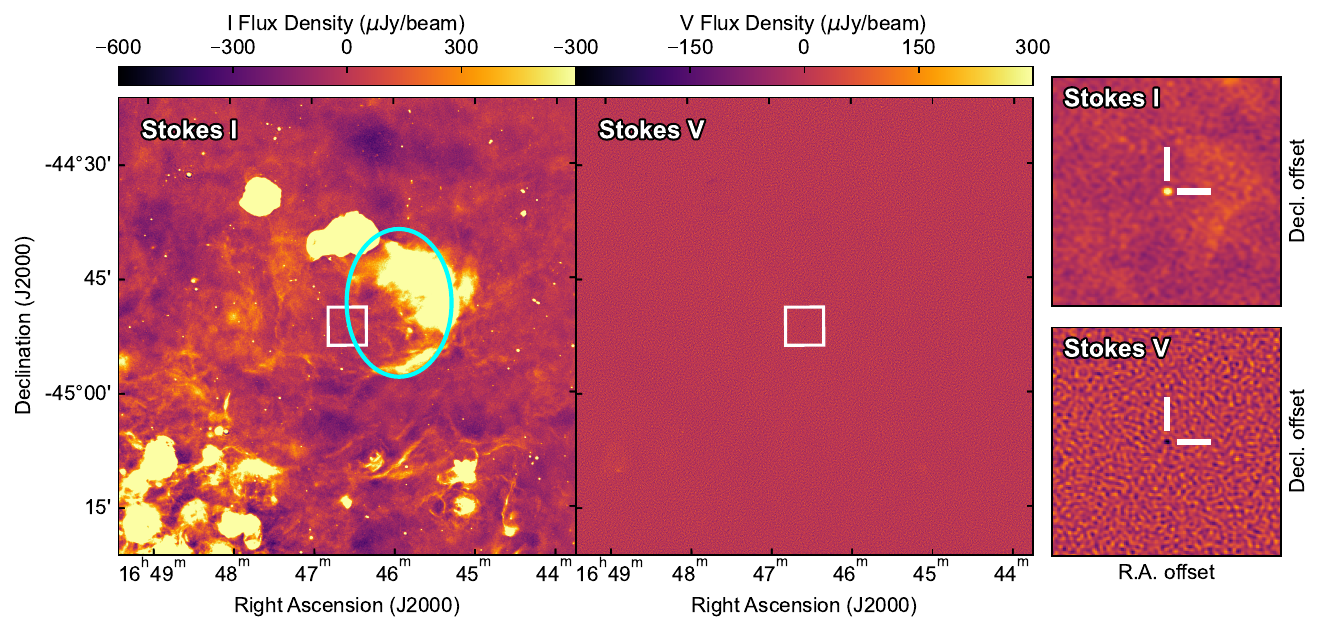}
    \includegraphics[width=0.95\textwidth]{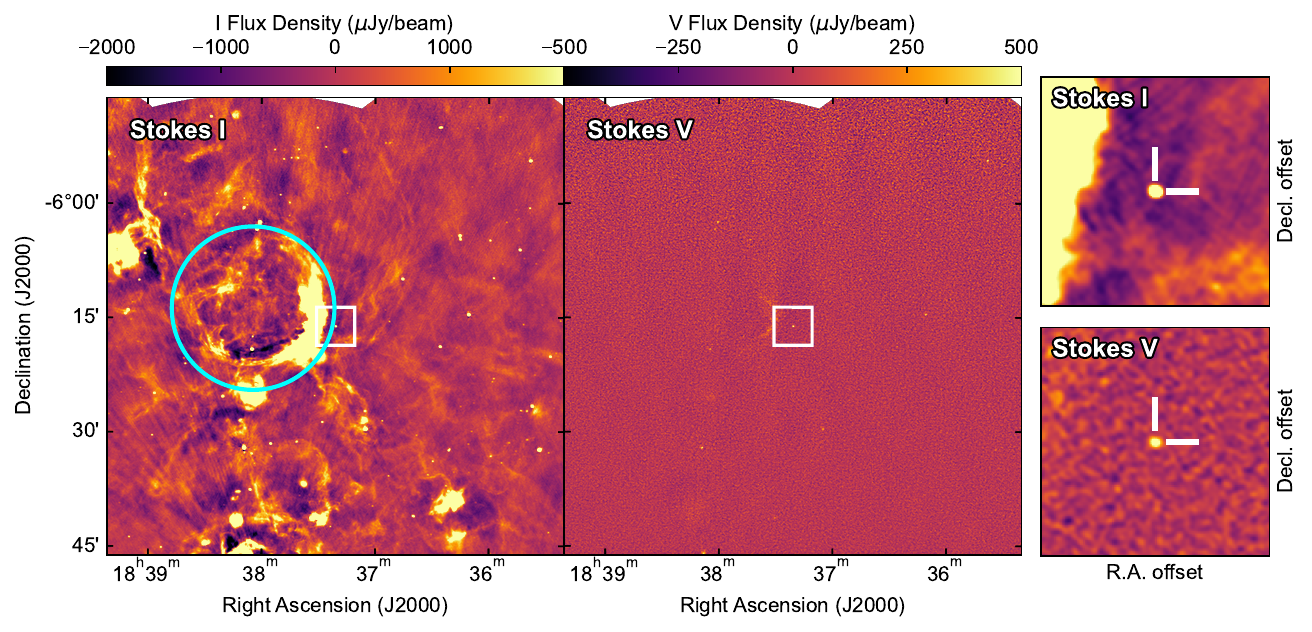}
    \caption{ASKAP radio continuum images of the fields centered on two pulsars at 888~MHz: PSR~J1646$-$4451 (top panels) and PSR~J1837$-$0616 (bottom panels). For each field, the Stokes~I image (left) shows the total intensity emission, revealing extended Galactic structure and diffuse background emission, while the Stokes~V image (middle) displays the circularly polarized intensity. The white squares mark the known positions of the pulsars. The panels on the right show zoomed-in views centered on the pulsar positions in both Stokes~I and V, with the pulsar location indicated by white crosshairs. These images have been corrected for the ASKAP primary beam response, and the color bars denote the flux density scales in $\mu$Jy\,beam$^{-1}$. Cyan ellipses indicate the \textsc{Hii} regions WISE\,G340.216$+$00.424 (top) and WISE\,G025.867$+$00.118 (bottom).}

    \label{fig:images}
\end{figure*}

To exploit the wide frequency coverage of the UWL band receiver, we implemented a subbanded search strategy. Each observation was divided into 12 frequency subbands--eight segments of 400~MHz and four segments of 800~MHz--resulting in 48 data files. Prior to searching, frequency masks for radio frequency interference (RFI) were generated using the \texttt{rfifind} suite from \textsc{PRESTO}\footnote{\url{https://github.com/scottransom/presto}} periodicity search code. Candidate selection following the acceleration search was performed according to the criteria established in \citet{sengar_23, sengar_htru_25}. Candidates were subsequently folded and diagnostic plots were generated using the \texttt{dspsr} and \texttt{pdmp} modules from the \textsc{PSRCHIVE}\footnote{\url{https://psrchive.sourceforge.net/}} software package. Folded profiles with a signal-to-noise ratio (S/N) exceeding 8 for slow pulsars ($P > 100$~ms) and 10 for fast pulsars ($P < 100$~ms) were shortlisted for visual inspection. Out of the four observations, three yielded clear detections of pulsars. To verify whether these were previously known sources, we used the Pulsar Survey Scraper\footnote{\url{https://pulsar.cgca-hub.org/}} to cross-check positional matches. Two of the pulsars, PSRs~J1646$-$4451 and J1837$-$0616, did not correspond to any published or cataloged sources and are therefore new pulsar detections (their corresponding radio continuum images are shown in Figure \ref{fig:images}).

Both new pulsars are extremely faint and exhibit significant scattering below 1500 MHz. PSR~J1646$-$4451 is particularly faint, and we initially detected it only in the 2400–3200 MHz and 3200–4000 MHz subbands with a S/N of approximately 13 in each subband. However, when the full 1500–4032 MHz band was coherently folded, the resulting folded S/N increased to 18. The full width at half maximum (FWHM) of the integrated pulse profile is approximately 10\% of the pulse period. In contrast, PSR~J1837$-$0616 was detected across all subbands above 1500 MHz. The folded S/N over the full 1500–4032 MHz band was 115, and the FWHM of its pulse profile is approximately 11\%.

\begin{deluxetable*}{l l l}
\tablecaption{Measured and derived timing parameters for PSR J1646$-$4451 and PSR J1837$-$0616 with 1--sigma uncertainties on the last digit quoted in parenthesis. Distances are estimated using both the YMW16 and NE2001 Galactic electron density models.\label{tab:timing}}
\tablehead{
\colhead{Parameter} & \colhead{PSR J1646$-$4451} & \colhead{PSR J1837$-$0616}}
\startdata
\tableline
\multicolumn{3}{c}{\text{Fitted parameters}}\\
\tableline
Right Ascension (J2000)\tablenotemark{a} & $16^{\rm h}46^{\rm m}34\fs66(1)$ & $18^{\rm h}37^{\rm m}20\fs916(6)$\\
Declination (J2000)\tablenotemark{a} & $-44{\degr}51\arcmin23.7(3)\arcsec$ & $-06{\degr}16\arcmin08.6(6)\arcsec$\\
Start (MJD) & 60240.0 & 60357.9\\
End (MJD) & 60581.1 & 60684.9\\
Number of TOAs & 22 & 74\\
Frequency (Hz) & 4.5973495725(6) & 8.4610639450(6)\\
Frequency Derivative (Hz\,s$^{-1}$) & $-4.574(4)\times10^{-14}$ & $-2.9991(5)\times10^{-13}$\\
Epoch of Period (MJD) & 60240.0 & 60391.9\\
$\chi^2$/DOF & 13/17 & 78/69\\
RMS residual ($\mu$s) & 313.7 & 292.3\\
Dispersion Measure (pc\,cm$^{-3}$) & 928(2) & 793(3)\\
Rotation Measure (rad\,m$^{-2}$) & $-134(7)$ & $1158(5)$\\
\tableline
\multicolumn{3}{c}{\text{Derived parameters}}\\
\tableline
Galactic Longitude (deg) & 340.251 & 25.749\\
Galactic Latitude (deg) & +0.305 & +0.261\\
Period (s) & 0.21751663305(2) & 0.1181884460(1)\\
Period Derivative (s\,s$^{-1}$) & $2.1643(1)\times10^{-15}$ & $4.1890(9)\times10^{-15}$\\
Characteristic Age (kyr) & 1593.4 & 447.3\\
Surface Magnetic Field (G) & $6.94\times10^{11}$ & $7.12\times10^{11}$\\
Spin-down Luminosity (erg\,s$^{-1}$) & $8.30\times10^{33}$ & $1.00\times10^{35}$\\
Distance (kpc) & 5.97\tablenotemark{b} & 5.75\tablenotemark{b}\\
 & 9.51\tablenotemark{c} & 8.78\tablenotemark{c}\\
\enddata

\tablenotetext{a}{Position from timing.}
\tablenotetext{b}{From the YMW16 \citep{ymw16} electron density model.}
\tablenotetext{c}{From the NE2001 \citep{ne2001} electron density model.}
\end{deluxetable*}

\section{Timing Analysis}\label{sec:timing}

\begin{figure}
    \includegraphics[width=0.99\columnwidth]{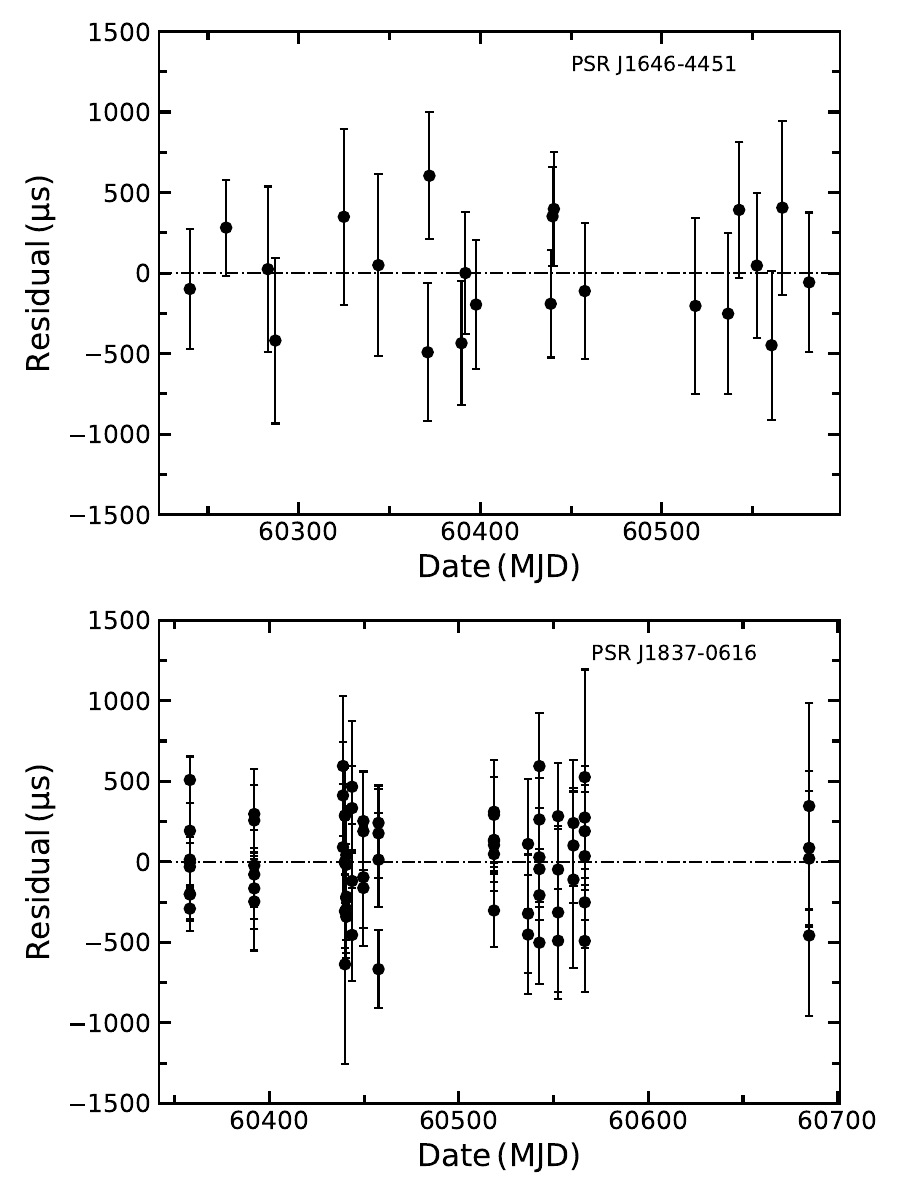}
    \caption{Timing residuals for PSR J1646$-$4451 and PSR J1837$-$0616 using Parkes data.}
    \label{fig:timing_residuals}
\end{figure}

\begin{figure*}
    \includegraphics[width=0.99\columnwidth]{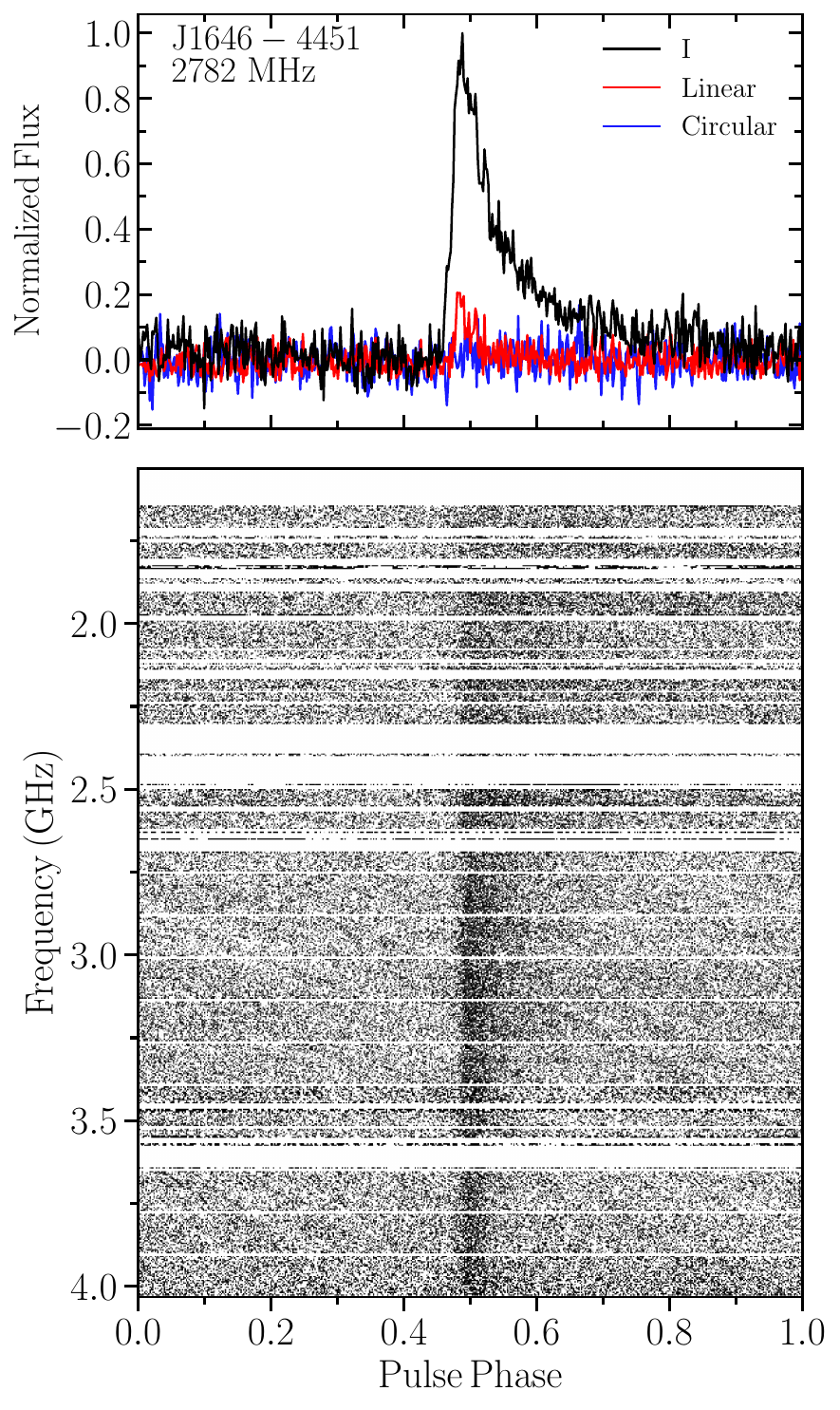}
    \includegraphics[width=0.99\columnwidth]{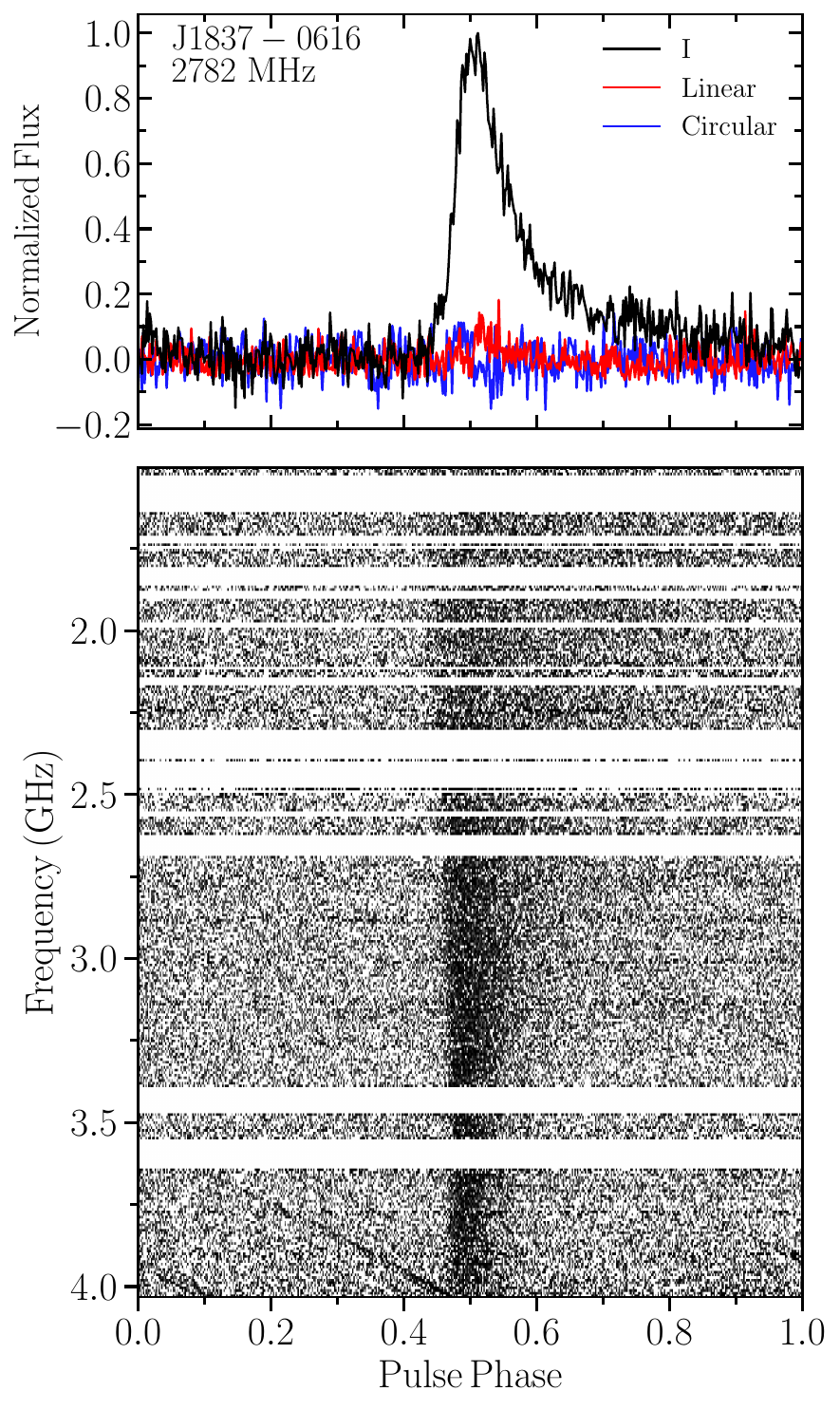}
    \caption{
        Pulse profiles and frequency-phase diagrams of PSR~J1646$-$4451 (left) and PSR~J1837$-$0616 (right) observed at a central frequency of 2782~MHz. \textit{Top panels:} Total intensity (Stokes $I$, black), linear polarization (red), and circular polarization (blue) as functions of pulse phase. The profiles are normalized to arbitrary units for comparative visualization. \textit{Bottom panels:} Frequency-resolved dynamic spectra (frequency vs.\ pulse phase) showing the evolution of pulse structure across the 1.5--4.0~GHz band. Persistent radio-frequency interference (RFI) has been masked in frequency.}
        
    \label{fig:freq vs phase}
\end{figure*} 

Following the discovery of the two pulsars, we conducted a follow-up campaign under the Parkes project P1332, using the UWL receiver in conjunction with the \textsc{Medusa} backend. Observations were conducted in search mode with full Stokes parameters along with a noise diode observation before observing each pulsar. Data were recorded with a frequency resolution of 1\,MHz per channel, yielding 3328 frequency channels, and a time resolution of 128\,$\mu$s.

PSR~J1646$-$4451 was discovered with a modest S/N ($\approx$ 20) during a 72-minute observation. Consequently, subsequent follow-up observations of this source were integrated over durations ranging from 60 to 90 minutes to maintain similar sensitivity in each observation. However, PSR~J1837$-$0616, being intrinsically brighter, required only 5–10 minutes of integration to achieve high-quality detections.

Each observation was initially folded at the known spin period and dispersion measure (DM) using \textsc{dspsr}, and the folding parameters were subsequently refined using \textsc{pdmp}. The resulting folded archives were cleaned using the \textsc{clfd}\footnote{\url{https://github.com/v-morello/clfd}} software to mitigate RFI. Additionally, narrowband and impulsive RFI were excised manually using \textsc{pazi}, part of the \textsc{PSRCHIVE} suite. Note that both pulsars exhibit strong scattering below 1500\,MHz and are almost invisible below 1500 MHz. Therefore, we excluded the 704–1500\,MHz portion of the UWL band from all subsequent analyses.

The highest S/N observations were selected to construct a standard template profile for each pulsar. Using these templates, time-of-arrival (TOA) measurements were generated with the \textsc{pat} utility from the \textsc{PSRCHIVE} package \citep{psrchive_12} by employing the Fourier-domain Markov Chain Monte Carlo (FDM) algorithm. Also, before creating the TOAs, the archive files were fully scrunched in frequency and polarization and in time they were scrunched such that each resulting TOA should have at least a minimum folded S/N of 15. For PSR~J1646$-$4451, one TOA was extracted per observation, yielding 22 TOAs across 22 epochs. For PSR~J1837$-$0616, multiple TOAs were obtained per observation, resulting in a total of 74 TOAs across 15 epochs.

Time-of-arrival (TOA) measurements were used to derive phase-connected timing solutions using the Pulsar INstrumentation and Timing \citep[\textsc{PINT};][]{pint_21} software package. All TOAs were corrected to the observatory clock and referenced to the solar system barycenter using the DE440 planetary ephemeris. As the pulsar positions determined from imaging are highly precise, we initially fit only for the spin frequency and its derivative, holding the position fixed at the imaging coordinates. However, we found that fitting for the position as well yielded improved timing solutions. The changes in position after fitting were sub-arcsecond for both pulsars, consistent with the imaging-derived positions.
Given the relatively good S/N of all TOAs, no TOAs were excluded from the analysis. We independently verified our timing solutions using TEMPO2 \citep{hobbs_06}, and obtained consistent results, which confirms the robustness of the solutions for both pulsars. The timing residuals are presented in Figure \ref{fig:timing_residuals}, and the corresponding timing parameters are summarized in Table \ref{tab:timing}. The timing baselines span approximately one year for both pulsars (341 days for PSR~J1646$-$4451 and 327 days for PSR~J1837$-$0616), and we therefore conclude that further observations are unlikely to significantly improve the precision of the reported solutions.

The pulsars have characteristic ages of 1593 kyr and 447 kyr, respectively, indicating that they are not particularly young and are consistent with the normal pulsar population. The surface magnetic field strengths are approximately $7.0 \times 10^{11}\,\mathrm{G}$. The spin-down luminosities, $8.3 \times 10^{33}\,\mathrm{erg\,s^{-1}}$ for PSR~J1646$-$4451 and $1.0 \times 10^{35}\,\mathrm{erg\,s^{-1}}$ for PSR~J1837$-$0616, are also typical of the normal pulsar population. The DM-derived distances for both pulsars are $\sim$6 kpc using the YMW16 model, and $\sim$8.8–9.5 kpc using the NE2001 model. These estimates place both pulsars several kiloparsecs away in the Galactic plane. However, these distance estimates differ by nearly 40\%, highlighting the uncertainties in distance determinations from dispersion measures.

\begin{figure*}[t]
    \centering
    \includegraphics[width=0.95\textwidth]{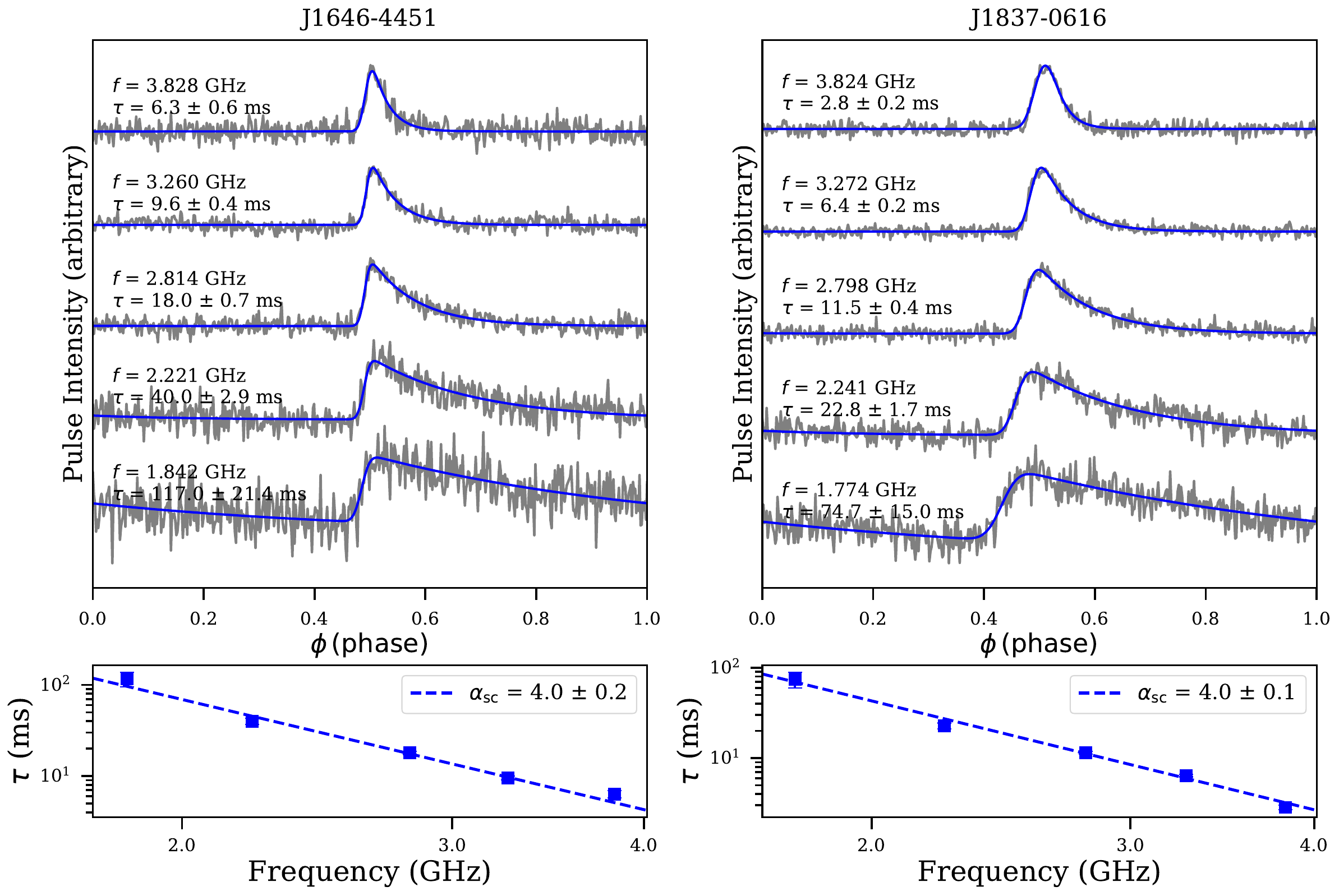}
    \caption{Pulse profiles of PSR J1646$-$4451 (left) and PSR J1837$-$0616 (right) at multiple observing frequencies. The grey curves in the upper panels show the observed pulse intensity as a function of rotational phase, while the blue curves represent best-fit models incorporating exponential scattering tails. Each sub-panel indicates the central observing frequency $f$ (in GHz) and the corresponding scattering timescale $\tau$ (in ms) with 1$\sigma$ uncertainties. The profiles exhibit clear frequency-dependent broadening due to interstellar scattering, with larger $\tau$ values at lower frequencies. The lower panels display the frequency dependence of the scattering timescales $\tau$ in log-log space for each pulsar. The blue dashed lines represent power-law fits of the form $\tau \propto \nu^{-\alpha_{\mathrm{sc}}}$, where $\alpha_{\mathrm{sc}}$ is the scattering index. }

    \label{fig:scattering_fit}
\end{figure*}

\subsection{Polarimetry}

We performed both flux and polarization calibration for PSR~J1646$-$4451 and PSR~J1837$-$0616 using standard calibration procedures. Each follow-up pulsar observation was preceded by a short-duration ($\sim$2-minute) noise diode observation, which served as a reference for flux and polarization calibration. All observations were first pre-processed for radio-frequency interference (RFI) excision using clfd and then remaining bad channels were manually removed using the \texttt{pazi} routine of PSRCHIVE.

For the flux density reference, we used archival observations of the bright, well-characterized radio quasar PKS~B1934$-$638. This source is regularly monitored as part of the Parkes Pulsar Timing Array project \citep[PPTA;][]{ppta_13} and provides a stable reference for both flux density and polarization calibration. The \texttt{pac} suite of the \texttt{PSRCHIVE} software package was employed to generate the calibrator solution database file from PKS~B1934$-$638 observations. These calibration solutions were then applied to the pulsar observations to obtain fully calibrated Stokes profiles. For each pulsar, individual calibrated observations were coherently coadded to produce a single high S/N profile, preserving both total intensity and polarization information. 

We subsequently used the \texttt{rmfit} utility within the \texttt{PSRCHIVE} suite to determine the rotation measure (RM) by fitting a Gaussian model to the RM spectrum derived from the frequency-resolved Stokes $Q$ and $U$ parameters. For PSR~J1646$-$4451, we obtained an RM detection of $-134 \pm 7$\,rad\,m$^{-2}$, which is typical for pulsars. For PSR~J1837$-$0616, we measured an RM of $1158 \pm 5$\,rad\,m$^{-2}$. The Stokes profiles and frequency-versus-phase plots for these pulsars are shown in Figure \ref{fig:freq vs phase}. Based on these pulse profiles, we found that PSR~J1646$-$4451 exhibits approximately 10\% linear polarization with no detectable circular polarization, whereas PSR~J1837$-$0616 shows negligible linear or circular polarization. This contrasts with the polarization properties observed in ASKAP imaging, where the circular polarization was found to be 19 \% and 7\%, respectively. 

The reason for this discrepancy is that the ASKAP's deep imaging continuum images are at 888\, MHz, in which the source is unresolved in the pulse phase and the Stokes parameters are averaged and preserved over the duration of the observation. By contrast, UWL detections are phase-resolved and are at higher effective frequencies, since these pulsars are so strongly scattered that they are effectively undetectable below $\sim$1.5\,GHz; the pulse profile and polarization are therefore measured only where the emission re-emerges at higher frequencies. In addition, the frequency dependence of circular polarization is highly diverse and can exhibit strong intrinsic frequency evolution and may either increase, decrease, or remain nearly constant with observing frequency as seen in many pulsars across the GHz band \citep[e.g.,][]{you_06}. In our case, it is possible that the intrinsic circular polarization fraction is present at $\sim$1 GHz and consistent with the ASKAP detection, but decreases toward higher frequencies such that, by $\gtrsim$1.5 GHz, the underlying circular polarization is already too weak and cannot be further differentiated from noise. On the other hand, interstellar scattering further reduces the significance of phase-resolved polarization by spreading pulsed power over a broader range of pulse phase while leaving the noise level unchanged, making any remaining circular component even harder to detect in pulsar profiles. Thus, a combination of intrinsic circular polarization  evolution with frequency and strong scattering provides a plausible explanation for circular polarization in the 888 MHz continuum images but negligible circular polarization in the phase-resolved UWL data.

\subsection{Scattering Analysis }

Using the coadded observations, we also investigated scattering characteristics of each source across frequency. For this purpose, we divided the coadded observations into several frequency subbands depending on their S/N. For PSR~J1646$-$4451, the total S/N of the coadded profile was 76. We divided the data into three subbands of 844\,MHz each, centered at 1920\,MHz, 2765\,MHz, and 3609\,MHz. For PSR~J1837$-$0616, the coadded S/N was 150, and we used four subbands of 633\,MHz bandwidth, centered at 1810\,MHz, 2450\,MHz, 3080\,MHz, and 3710\,MHz. We fitted an exponential pulse broadening function convolved with a Gaussian to each subband profile to extract the scattering timescale \( \tau_{\mathrm{sc}} \) at each frequency of the form:
\[
I(\phi) = I_0 \cdot \exp\left(-\frac{\phi - \phi_0}{\tau_{\mathrm{sc}}}\right), \quad \phi > \phi_0,
\]
where $I(\phi)$ represents the intensity as a function of pulse phase $\phi$, $I_0$ is the amplitude at reference phase $\phi_0$, and $\tau_{\mathrm{sc}}$ denotes the scattering timescale characterizing pulse broadening from multipath propagation through the interstellar medium. This model attributes the observed scattering tail to path-length differences introduced by small-scale electron-density inhomogeneities in the ionized interstellar medium while assuming an intrinsic Gaussian pulse shape. The resulting frequency-dependent scattering timescales for both pulsars are presented in Figure \ref{fig:scattering_fit}.

For PSR~J1646$-$4451, the scattering timescale measured from the scattered profile at the lowest observing frequency, 1.842\,GHz, is $117 \pm 21$\,ms. For PSR~J1837$-$0616, we measure a scattering timescale of $74.7 \pm 15$\,ms at 1.774\,GHz. We model the frequency dependence of the scattering timescale using a power-law:

\[
\tau_{\mathrm{sc}}(\nu) = \tau_0 \left( \frac{\nu}{1\,\mathrm{GHz}} \right)^{-\alpha},
\]
where \( \tau_0 \) is the scattering timescale at 1\,GHz and \( \alpha \) is the scattering index. A weighted least-squares fit yielded the same scattering index for both pulsars, i.e., \( \alpha = 4.0 \pm 0.2 \) for PSR~J1646$-$4451 and \( \alpha = 4.0 \pm 0.1 \) for PSR~J1837$-$0616. These values are somewhat flatter than the Kolmogorov turbulence prediction of  \citep[ $\alpha \sim 4.4 $;][]{lee76}, however, they are consistent with the thin screen scattering model \citep[e.g.][]{cronym_70, rickett_2000}.

\begin{figure}
    \includegraphics[width=0.95\columnwidth]{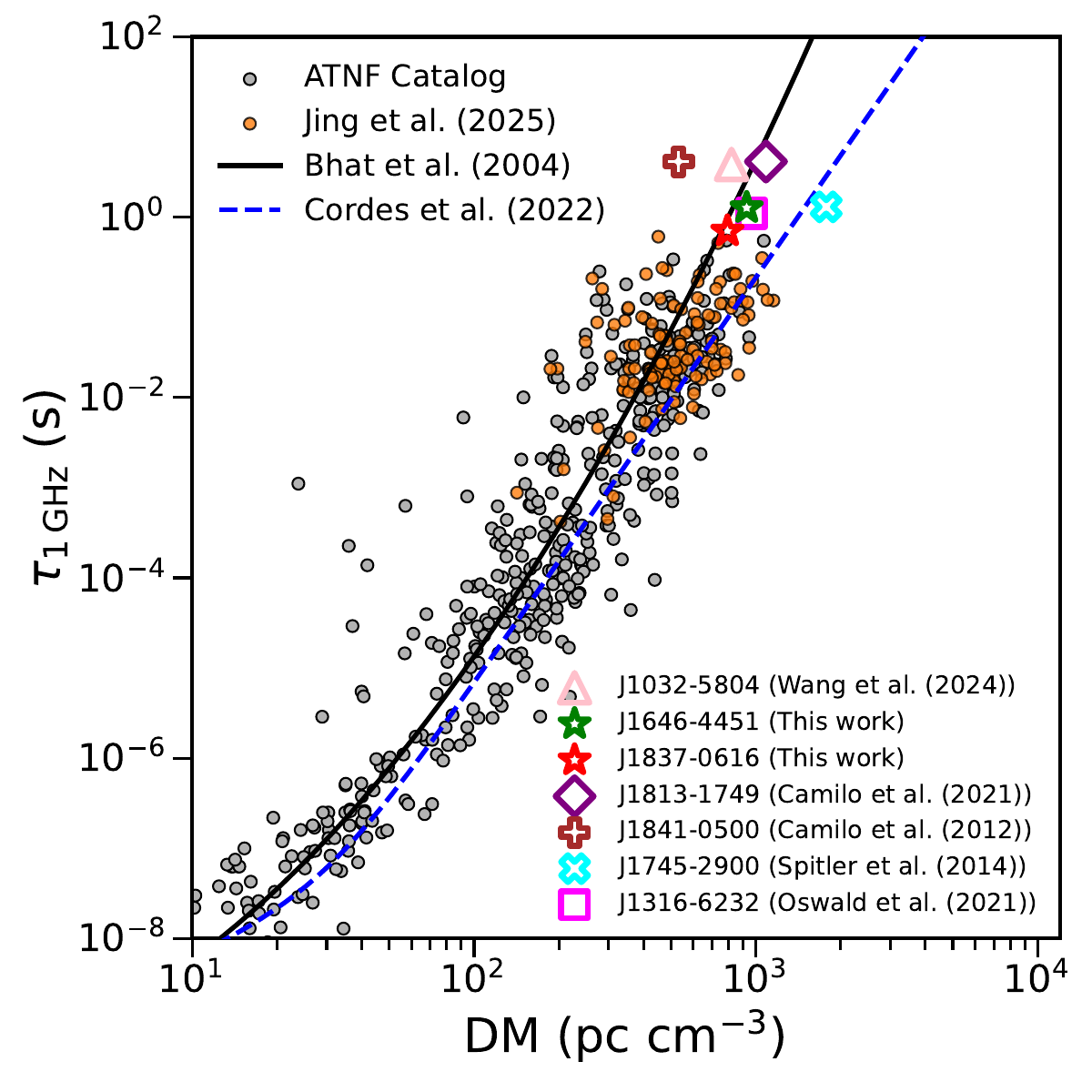}
    \caption{ Measured scattering timescales at 1 GHz $\tau_{\rm sc, 1\mathrm{GHz}}$ as a function of dispersion measure (DM) for a sample of pulsars. Grey circles represent archival measurements from the ATNF pulsar catalogue \citep[PSRCAT;][]{manchester_05} version 2.7, and orange circles correspond to the published $\tau_{1\,\mathrm{GHz}}$ of pulsars discovered in FAST GPPS survey \citep{jing_2025}. The solid black line and dashed blue line denote the empirical relations from \citet{bhat_04} and \citep{cordes_22}, showing the expected dependence of scattering on DM. The green and red star mark the locations of PSR~J1646$-$4451 and PSR~J1837$-$0616, respectively, as determined from our scattering analysis. Both sources lie near the upper envelope of the distribution, indicating relatively strong scattering along their lines of sight.}
    \label{fig:tau_vs_dm}
\end{figure}

Using the measured scattering indices, the extrapolated scattering timescales at 1\,GHz, $\tau_{\rm sc,1\,GHz}$ are 1346(296) ms for PSR~J1646$-$4451 and 740(154) ms for PSR~J1837$-$0616. These values indicate that both pulsars are among the most highly scattered pulsars known, and consistent with expectations for sources located along dense or turbulent lines of sight in the Galactic plane. These scattering timescale values are plotted in Figure~\ref{fig:tau_vs_dm}, along with the empirical relation from \citet{bhat_04} and \citet{cordes_22}. 
The scattering timescale for PSR~J1837$-$0616 is in good agreement with the \citet{bhat_04} model, differing by $\sim$30\%. In contrast, the timescale for PSR~J1646$-$4451 at 1\,GHz is approximately a factor of two lower than the predicted value which is expected given the substantial intrinsic scatter and uncertainties inherent in such empirical models.

\section{Discussion} \label{sec:discussion}

No new pulsations were detected in the periodicity searches of the remaining two circularly polarized sources.   One of those is likely associated with a star.  The nature of the other source, J1832$-$0808, is less clear.  However, one known pulsar, PSR~B1829$-$08, was detected in the observation of that source. PSR~B1829$-$08 is a bright pulsar with a well-determined position and timing solution. It lies approximately $19\arcmin$ from the beam center. Given both its brightness and the size of the Parkes beam ($\approx 28\arcmin$), it is likely that this detection is due to a sidelobe response, rather than a physical association with the circularly polarized source. Bright pulsars like B1829$-$08 can also often be detected at relatively large off-axis angles.

\begin{figure*}[t]
    \includegraphics[width=1.99\columnwidth]{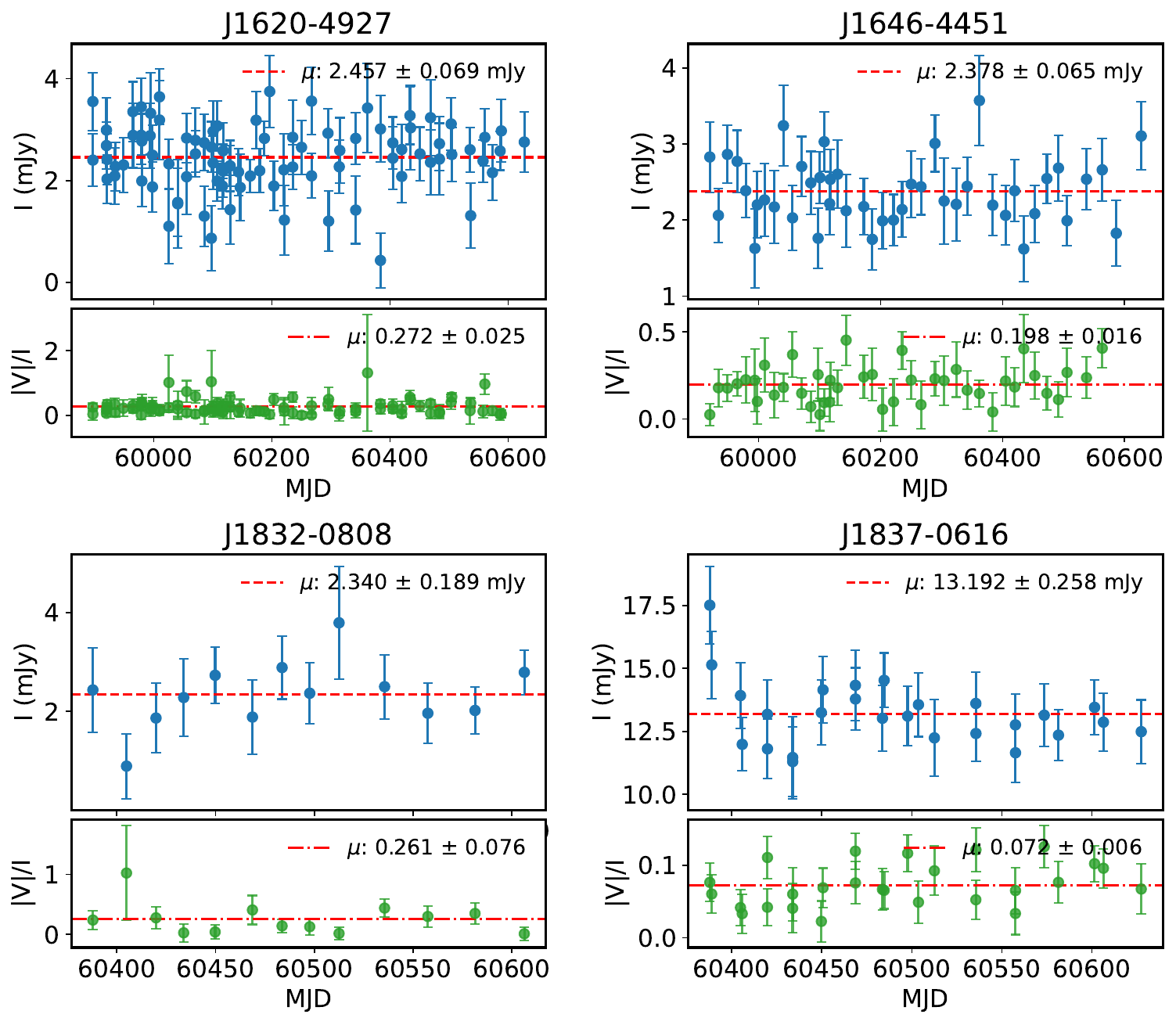}
    \caption{Total intensity ($I$, top panels) and circular polarization ($V$, bottom panels) as a function of MJD for four circularly polarized sources. Each point represents a single observation, with error bars denoting measurement uncertainties. The red dashed lines correspond to the mean flux densities in $I$ and $V$. The sources exhibit stable fluxes and polarization over time, with little variation.}
    \label{fig:light_curves}
\end{figure*}

The apparent lack of pulsations in \srcc, despite its continuum properties being similar to those of pulsars, motivated us to investigate whether the pulsations could be missed somehow.  Although it is almost an order of magnitude higher than the average circular polarization observed in pulsars, it is consistent with circular polarization values detected in the known pulsar population in the Rapid ASKAP Continuum Survey \citep{anumarlapudi_23}. Its mean Stokes~$I$ flux density is approximately 2.4\,mJy at 843--943.5\,MHz. Scaling the flux density to 1.4\,GHz, and assuming a steep spectral index of $-2.0$, its expected flux density at 1.4\,GHz would be $\sim1.0$\,mJy. Furthermore, based on the available ASKAP epochs, its total intensity and circular polarization appear stable over time, with no strong evidence of variability (see Figure \ref{fig:light_curves}). These properties are broadly consistent with typical radio pulsars. However, the mean S/N for this source is relatively low as compared to the confirmed pulsars in the sample. One plausible explanation for the non-detection of this source in follow-up searches is that it is simply too faint to be detected in the  $\sim$1\,hour Parkes observations.

\begin{figure}
    \includegraphics[width=0.99\columnwidth]{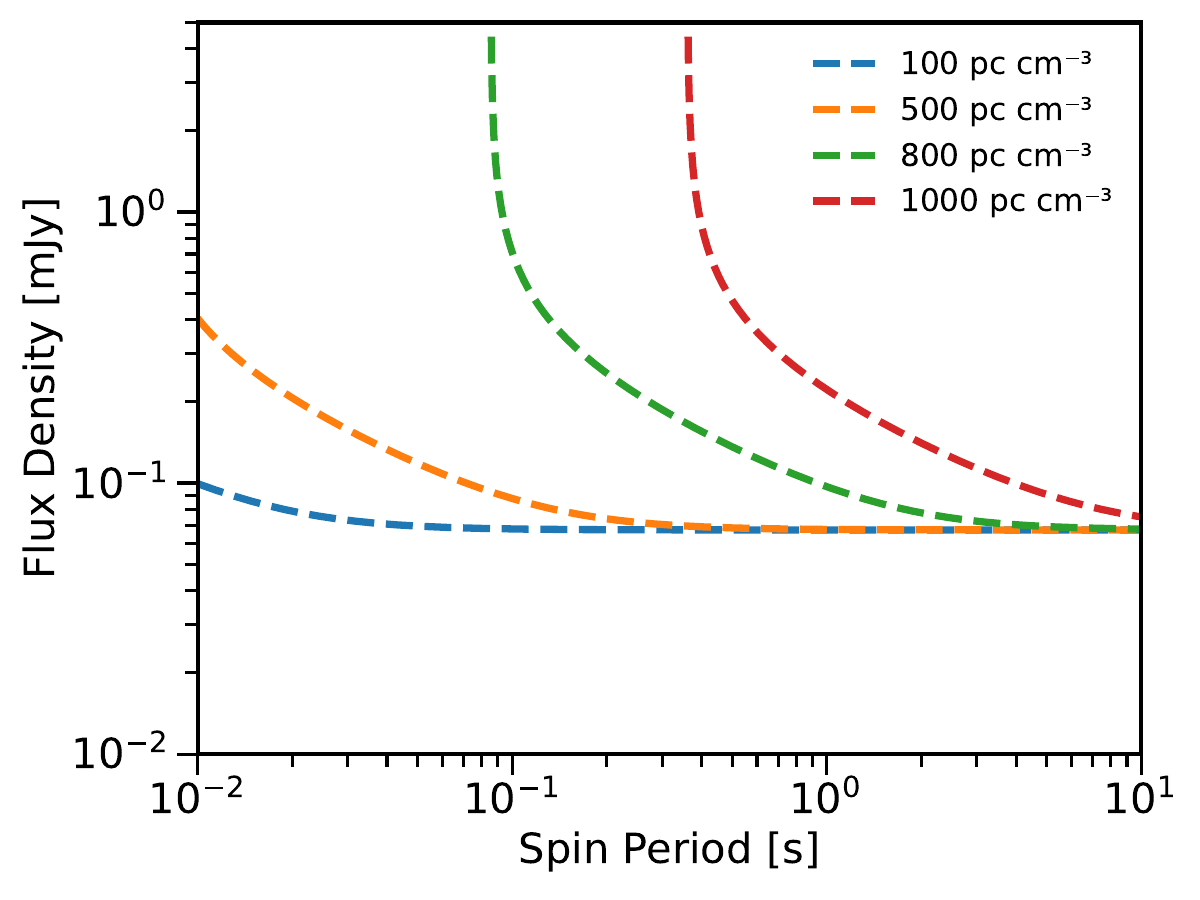}
    \caption{Minimum detectable flux density as a function of pulsar spin period for various dispersion measures (DMs), assuming a fixed Parkes observing setup with a central frequency of 1400~MHz, bandwidth of 500~MHz, and an integration time of 1~hour. Sensitivity curves are shown for DMs of 100, 500, 800, and 1000~pc~cm$^{-3}$. The effective pulse width used in computing these curves is given by $W_{\mathrm{eff}} = \sqrt{W_{\mathrm{int}}^2 + \tau_{\mathrm{scat}}^2}$, where $W_{\mathrm{int}}$ is the intrinsic pulse width, corresponds to a 5\% duty cycle, and $\tau_{\mathrm{scat}}$ is the scattering timescale estimated for the DMs using the empirical relation from \citet{bhat_04}. At higher DMs, sensitivity to fast pulsars is significantly degraded due to dispersion and scattering effects, illustrating the observational bias against detecting short-period pulsars in high-DM environments.}
    \label{fig:sensitivity_curves}
\end{figure}

To evaluate this, we considered the flux-density sensitivity curves for the Parkes telescope at L-band (center frequency: 1250~MHz, bandwidth: 500~MHz) for a 1.2-hour integration. In the absence of scattering, pulsars with spin periods $>100$\,ms, flux densities $\gtrsim 0.1$\,mJy, and DMs $\lesssim 1500$\,cm$^{-3} \, \rm pc$ should be detectable. However, when accounting for scattering using the empirical relation from \citet{bhat_04}, pulsars with periods $<200$~ms, DMs $>800$\,cm$^{-3} \, \rm pc$, and flux densities $\lesssim 0.2$\,mJy may fall below the detection threshold due to significant pulse broadening at L-band (see Figure \ref{fig:sensitivity_curves}). Therefore, if  J1832$-$0808 is indeed a pulsar, its non-detection could be attributed to several factors: it may be fast-spinning ($P < 100$~ms), lie at high DM ($>800$~cm$^{-3} \, \rm pc$), and suffer from severe scattering, rendering it undetectable at L-band. Unlike the two confirmed pulsars, this source may be intrinsically faint and possess steeper radio spectra. Given these considerations, future searches with high-sensitivity telescopes such as MeerKAT would be particularly valuable for confirming the nature of these sources.

\begin{figure}
    \centering
    \includegraphics[width=0.99\columnwidth]{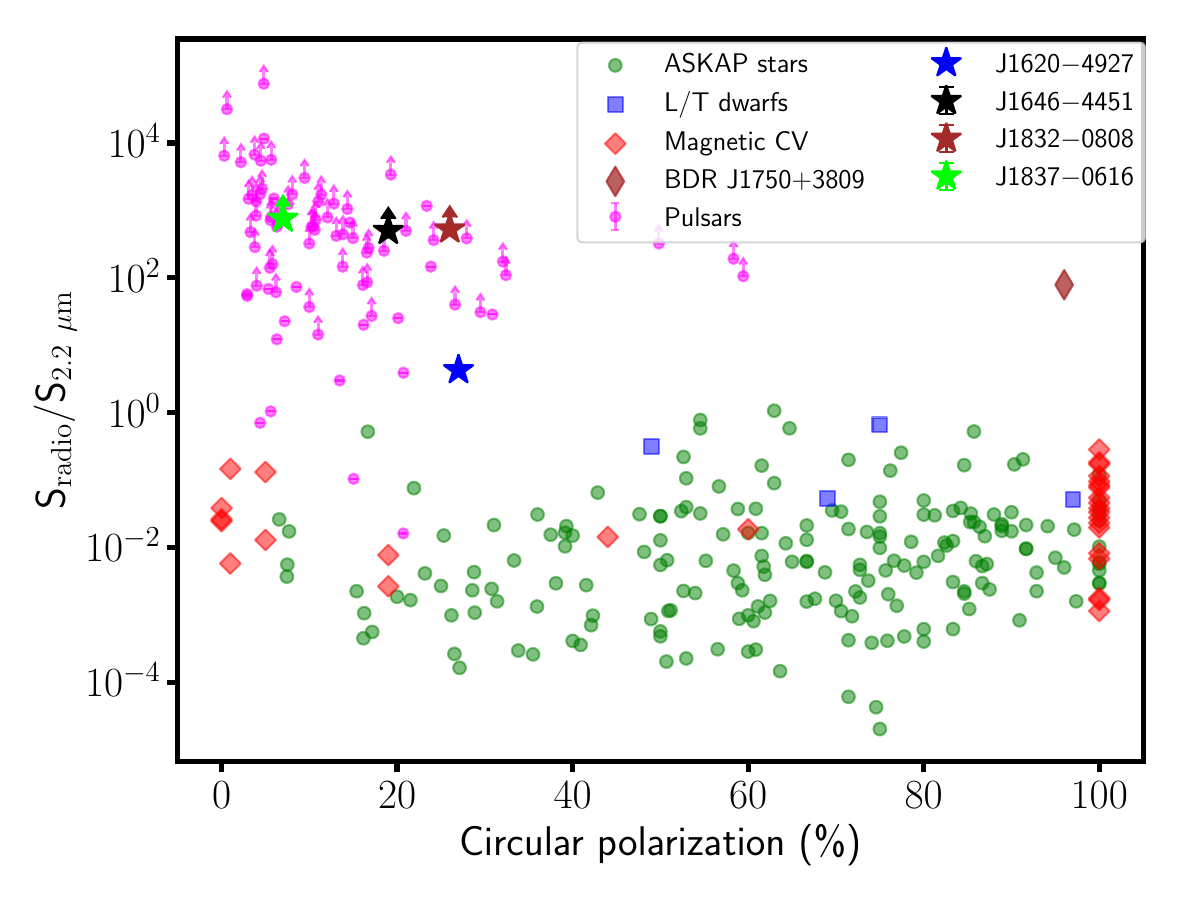}

    \caption{Phase space of circular polarization fraction and the relative source strength in radio and $K_s$ bands showing different populations of sources (also see \citealt{wang_22}). The green circles represent radio-detected stars \citep{driessen_24}, the red diamonds represent MCVs \citep{barrett2020}, the blue squares show dwarf stars \citep{pritchard_21}, and the pink circles represent pulsars. Individual sources of interest, including the radio-discovered brown dwarf BDR J0750+3809 \citep{Vedantham2020} and our sample, are highlighted.}
    \label{fig:pol_vs_radio_nir}
\end{figure}

Given the optical/NIR+radio analysis presented above, three of which resulted in non-detections (or accidental associations), we adopted a phenomenological approach to understand the nature of \srcc\ by comparing their radio and $K_s$-band properties. Figure~\ref{fig:pol_vs_radio_nir} shows different source classes that emit significantly polarized radio emission, in the phase space of their polarization properties and the relative radio to $K_s$-band continuum strength. It can be seen that \srca, a radio star, has a similar radio-to-NIR flux ratio that is a factor of $>10$ more extreme than most stars selected from a circular polarization search \citep{pritchard_21}, but it is consistent with the radio-discovered brown dwarf BDR J0750+3809 \citep{Vedantham2020} even though they have very different spectral types and polarization levels.  This may suggest that stars when identified through radio discovery may span a wider range of properties than previously thought. The other three sources in our sample lie in a region occupied by pulsars, suggesting that \srcc\ may indeed be a pulsar despite the lack of detected pulsations. 

\section{Conclusion} \label{sec:conclusion}

We report the discovery of two highly scattered pulsars, PSRs J1646$-$4451 and J1837$-$0616, identified through their circular polarization in ASKAP continuum images and confirmed via targeted observations with the Parkes Ultra-Wideband Low (UWL) receiver. Both PSR J1646$-$4451 and PSR J1837$-$0616 exhibit spin characteristics that are typical of the normal pulsar population. Their spin periods—217 ms for PSR J1646$-$4451 and 118 ms for PSR J1837$-$0616, place them within the region of the $P$–$\dot{P}$ diagram that is commonly occupied by young pulsars, including those associated with supernova remnants. However, their period derivatives are consistent with the broader population of normal, non-recycled pulsars. PSR J1646$-$4451 has a characteristic age of approximately 1.59 Myr, while PSR J1837$-$0616 is relatively younger, with a characteristic age of about 0.44 Myr. Despite the latter's lower age, it does not fall into the category of truly young pulsars, which are typically much younger ($< 50\,\rm{kyr}$) and often show direct associations with supernova remnants.

In addition to their typical spin properties, both pulsars exhibit high dispersion measures (DMs) and scattering characteristics. For both pulsars, their frequency-dependent scattering timescales (see Figure~\ref{fig:scattering_fit}) yield scattering indices of $\sim$4.0, which is consistent with the thin screen scattering model. The $\tau_{\rm sc,1\,GHz}$ values for PSR~J1646$-$4451 and PSR~J1837$-$0616 are 1346~ms and 740~ms, respectively, placing them among the seven most heavily scattered pulsars known.
Among these seven, PSR~J1813$-$1749 was first detected in X-rays within the supernova remnant G12.82$-$0.02 and later confirmed through high-frequency radio observations, with $\tau_{\rm sc, 1\,GHz}=4.14$~s \citep{camilo_21}. PSR~J1841$-$0500 \citep{camilo_11} has a comparable value of $\tau_{\rm sc, 1\,GHz}=4.12$~s. PSR J1316$-$6232 and the Galactic-center magnetar PSR~J1745$-$2900 both have similar $\tau_{\rm sc, 1\,GHz}$ of 1.1 and 1.3~s, respectively. Notably, the remaining three pulsars--including the two reported in this work and PSR~J1032$-$5804--were all initially identified as compact, circularly polarized sources in ASKAP imaging.
This emerging trend strongly suggests that identifying pulsar candidates using image-domain polarization search techniques and confirming them using wideband receivers serves as a powerful diagnostic for uncovering unusual pulsars particularly in high-DM, inner-Galactic regions where scattering is most severe.

Many of the sources identified through this method may represent a population of pulsars that are intrinsically young, rapidly spinning, or members of binary systems—categories that are often underrepresented in traditional pulsar surveys due to extreme scattering and associated selection biases. The discovery of two pulsars from four candidates (a 50\% success rate) further validates circular polarization as an efficient discriminant for pulsar searches in imaging surveys. Traditional periodicity searches at $\nu < 2$\,GHz would likely have missed these sources due to scattering smearing exceeding their spin periods. This approach complements time-domain surveys by probing populations in regions of high interstellar medium (ISM) turbulence, where $\tau_{\rm sc} \propto \nu^{-\alpha}$ suppresses low-frequency detections. Similar strategies have uncovered other highly scattered pulsars (e.g., PSR J1638$-$4713; \citealt{Lazarevi_24}), underscoring the necessity of wideband receivers like UWL to detect high-frequency signals above the scattering horizon.

The severe scattering observed towards these pulsars may be attributed to their locations relative to nearby \textsc{Hii} regions. Both \srcb{} and \srcd{} lie on the edge of large Galactic \textsc{Hii} regions, as seen by the Widefield Infrared Survey Explorer \citep[WISE;][]{2010AJ....140.1868W}. The two regions are seen in Figure~\ref{fig:images} as $\sim20'$-diameter, shell-like structures; respectively these pulsars are closest to WISE\,G340.216$+$00.424 and WISE\,G025.867$+$00.118. The pulsars are apparently coincident with the edge of the 12\,$\mu$m emission derived from the UV fluorescence of polycyclic aromatic hydrocarbons typically found around \textsc{Hii} regions. No reliable distance estimates are available in the literature for these sources, although we may make an estimate based on their size. The median \textsc{Hii} region diameter in the WISE \textsc{Hii} catalogue \citep{2014ApJS..212....1A} is $\sim3$\,pc, which would place these objects $\sim500$\,pc away. This is consistent with the low-frequency (free-free) absorption signal seen in the GaLactic and Extragalactic All-sky Murchison Widefield Array \citep[GLEAM;][]{2015PASA...32...25W,2019PASA...36...47H} and GLEAM-eXtended \citep[GLEAM-X;][]{2022PASA...39...35H,2025PASA...42..137M} data toward these objects.

\textsc{Hii} regions have been seen to increase the level of scattering for pulsars detected behind them, along the line-of-sight \citep[e.g.][]{1984A&A...135..199D}. A similar effect could be at play here, and would be maximised if these pulsars lie at twice the distance of the intervening \textsc{Hii} regions. Careful modelling and high-quality data on both the pulsars and the \textsc{Hii} region can be used to solve for the intrinsic properties of the sources \citep[see e.g.][]{2017MNRAS.467.3642S} but this lies beyond the scope of this paper.

Current and future large-scale radio imaging projects, such as an ongoing EMU survey with ASKAP and surveys with upcoming SKA, and DSA-2000 telescopes, are poised to substantially broaden the parameter space for pulsar discoveries by leveraging radio imaging. These capabilities will be especially valuable for detecting pulsars obscured by scattering in the Galactic plane, which remain challenging to traditional time-domain searches. Crucially, wideband follow-up observations will play an instrumental role in confirming such candidates and characterizing their properties, particularly for pulsars embedded in dense environments. Very recently, selecting pulsar candidates via circular polarization in radio continuum images \citep{frail_24} and then following them up with the Parkes UWL receiver has led to the discovery of several MSPs \citep{sengar_gb_26a}, and demonstrates how the combination of radio imaging and targeted wideband follow-up can uncover pulsars that are difficult to detect in traditional pulsar surveys.

Collectively, these developments highlight the complementary strengths of imaging and time-domain methodologies in uncovering diverse neutron star populations. Polarized continuum surveys, in particular, are emerging as a powerful tool for completing the Galactic pulsar inventory and probing small-scale structure in the turbulent interstellar medium. At the same time, these advances underscore the continued need for improved scattering models of radio wave propagation through highly scattered regions.

\begin{acknowledgments}

We thank our referee, Joris Verbiest, for the helpful comments that improved the manuscript. R.S.\ is supported by NSF grant AST-1816904 and grateful for the continuing support of the Max Planck Society. D.K.\ is supported by NSF grants AST-1816492, AST-1816904, and AST-2511757. This scientific work uses data obtained from Inyarrimanha Ilgari Bundara/the Murchison Radio-astronomy Observatory. N.H.-W. is the recipient of an Australian Research Council Future Fellowship (project number FT190100231). We acknowledge the Wajarri Yamaji People as the Traditional Owners and native title holders of the Observatory site. CSIRO’s ASKAP radio telescope is part of the Australia Telescope National Facility (\url{https://ror.org/05qajvd42}). Operation of ASKAP is funded by the Australian Government with support from the National Collaborative Research Infrastructure Strategy. ASKAP uses the resources of the Pawsey Supercomputing Research Centre. Establishment of ASKAP, Inyarrimanha Ilgari Bundara, the CSIRO Murchison Radio-astronomy Observatory, and the Pawsey Supercomputing Research Centre are initiatives of the Australian Government, with support from the Government of Western Australia and the Science and Industry Endowment Fund. The Parkes radio telescope is part of the Australia Telescope National Facility, funded by the Australian Government for operation as a National Facility managed by CSIRO. We acknowledge the Wiradjuri People as the Traditional Owners of the Parkes Observatory site. The Australia Telescope Compact Array is also part of the Australia Telescope National Facility (\url{https://ror.org/05qajvd42}), and we acknowledge the Gomeroi People as the Traditional Owners of its site. Parts of this research were supported by the Australian Research Council Centre of Excellence for Gravitational Wave Discovery (OzGrav), project number CE230100016.
\end{acknowledgments}

\vspace{5mm}
\facilities{ASKAP, Parkes, WISE (\url{https://www.ipac.caltech.edu/doi/irsa/10.26131/IRSA153}).}

\software{astropy \citep{2013A&A...558A..33A,2018AJ....156..123A},  PINT \citep{pint_21}, PSRCHIVE \citep{2004PASA...21..302H},  Pulsar Survey Scraper \citep{2022ascl.soft10001K}, PyGEDM \citep{2021PASA...38...38P}.}

\bibliography{vast_pulsars}{}

@ARTICLE{kaplan_19,
       author = {{Kaplan}, David L. and {Dai}, Shi and {Lenc}, Emil and {Zic}, Andrew and {Swiggum}, Joseph K. and {Murphy}, Tara and {Anderson}, Craig S. and {Cameron}, Andrew D. and {Dobie}, Dougal and {Hobbs}, George and {Kaczmarek}, Jane F. and {Lynch}, Christene and {Toomey}, Lawrence},
        title = "{Serendipitous Discovery of PSR J1431-6328 as a Highly Polarized Point Source with the Australian SKA Pathfinder}",
      journal = {\apj},
         year = 2019,
        month = oct,
       volume = {884},
       number = {1},
          eid = {96},
        pages = {96},
          doi = {10.3847/1538-4357/ab397f},
archivePrefix = {arXiv},
       eprint = {1908.03163},
 primaryClass = {astro-ph.HE},
       adsurl = {https://ui.adsabs.harvard.edu/abs/2019ApJ...884...96K}
}

@ARTICLE{sengar_gb_26a,
       author = {{Sengar}, Rahul and {Anumarlapudi}, Akash and {Kaplan}, David L. and {Frail}, Dale A. and {Hyman}, Scott D. and {Polisensky}, Emil},
        title = "{Millisecond Pulsar Discoveries in an Image-based MeerKAT Survey of the Galactic Bulge}",
      journal = {arXiv e-prints},
         year = 2025,
        month = sep,
          eid = {arXiv:2509.20614},
        pages = {arXiv:2509.20614},
          doi = {10.48550/arXiv.2509.20614},
archivePrefix = {arXiv},
       eprint = {2509.20614},
 primaryClass = {astro-ph.HE},
       adsurl = {https://ui.adsabs.harvard.edu/abs/2025arXiv250920614S}
}

@ARTICLE{cronym_70,
       author = {{Cronyn}, Willard M.},
        title = "{Interstellar Scattering of Pulsar Radiation and Its Effect on the Spectrum of NP0532}",
      journal = {Science},
         year = 1970,
        month = jun,
       volume = {168},
       number = {3938},
        pages = {1453-1455},
          doi = {10.1126/science.168.3938.1453},
       adsurl = {https://ui.adsabs.harvard.edu/abs/1970Sci...168.1453C}
}

@ARTICLE{rickett_2000,
       author = {{Rickett}, B.~J. and {Coles}, Wm. A. and {Markkanen}, Jussi},
        title = "{Interstellar Scintillation of Pulsar B0809+74}",
      journal = {\apj},
         year = 2000,
        month = apr,
       volume = {533},
       number = {1},
        pages = {304-319},
          doi = {10.1086/308637},
archivePrefix = {arXiv},
       eprint = {astro-ph/9911368},
 primaryClass = {astro-ph},
       adsurl = {https://ui.adsabs.harvard.edu/abs/2000ApJ...533..304R}
}

@ARTICLE{shimwell_17,
       author = {{Shimwell}, T.~W. and {R{\"o}ttgering}, H.~J.~A. and {Best}, P.~N. and {Williams}, W.~L. and {Dijkema}, T.~J. and {de Gasperin}, F. and {Hardcastle}, M.~J. and {Heald}, G.~H. and {Hoang}, D.~N. and {Horneffer}, A. and {Intema}, H. and {Mahony}, E.~K. and {Mandal}, S. and {Mechev}, A.~P. and {Morabito}, L. and {Oonk}, J.~B.~R. and {Rafferty}, D. and {Retana-Montenegro}, E. and {Sabater}, J. and {Tasse}, C. and {van Weeren}, R.~J. and {Br{\"u}ggen}, M. and {Brunetti}, G. and {Chy{\.z}y}, K.~T. and {Conway}, J.~E. and {Haverkorn}, M. and {Jackson}, N. and {Jarvis}, M.~J. and {McKean}, J.~P. and {Miley}, G.~K. and {Morganti}, R. and {White}, G.~J. and {Wise}, M.~W. and {van Bemmel}, I.~M. and {Beck}, R. and {Brienza}, M. and {Bonafede}, A. and {Calistro Rivera}, G. and {Cassano}, R. and {Clarke}, A.~O. and {Cseh}, D. and {Deller}, A. and {Drabent}, A. and {van Driel}, W. and {Engels}, D. and {Falcke}, H. and {Ferrari}, C. and {Fr{\"o}hlich}, S. and {Garrett}, M.~A. and {Harwood}, J.~J. and {Heesen}, V. and {Hoeft}, M. and {Horellou}, C. and {Israel}, F.~P. and {Kapi{\'n}ska}, A.~D. and {Kunert-Bajraszewska}, M. and {McKay}, D.~J. and {Mohan}, N.~R. and {Orr{\'u}}, E. and {Pizzo}, R.~F. and {Prandoni}, I. and {Schwarz}, D.~J. and {Shulevski}, A. and {Sipior}, M. and {Smith}, D.~J.~B. and {Sridhar}, S.~S. and {Steinmetz}, M. and {Stroe}, A. and {Varenius}, E. and {van der Werf}, P.~P. and {Zensus}, J.~A. and {Zwart}, J.~T.~L.},
        title = "{The LOFAR Two-metre Sky Survey. I. Survey description and preliminary data release}",
      journal = {\aap},
         year = 2017,
        month = feb,
       volume = {598},
          eid = {A104},
        pages = {A104},
          doi = {10.1051/0004-6361/201629313},
archivePrefix = {arXiv},
       eprint = {1611.02700},
 primaryClass = {astro-ph.IM},
       adsurl = {https://ui.adsabs.harvard.edu/abs/2017A&A...598A.104S}
}

@ARTICLE{hopkins_25,
       author = {{Hopkins}, Andrew and {Kapinska}, Anna and {Marvil}, Joshua and {Vernstrom}, Tessa and {Collier}, Jordan and {Norris}, Ray and {Gordon}, Yjan and {Duchesne}, Stefan and {Rudnick}, Lawrence and {Gupta}, Nikhel and {Carretti}, Ettore and {Anderson}, Craig and {Dai}, Shi and {G{\"u}rkan}, Gulay and {Parkinson}, David and {Prandoni}, Isabella and {Riggi}, Simone and {Shekhar Saraf}, Chandra and {Ma}, Yik Ki and {Filipovi{\'c}}, Miroslav D. and {Umana}, Grazia and {Bahr-Kalus}, Benedict and {Koribalski}, B{\"a}rbel Silvia and {Lenc}, Emil and {Ingallinera}, Adriano and {Afonso}, Jos{\'e} and {Ahmad}, Adeel and {Ahmed}, Ummee Tania and {Alexander}, Emma and {Andernach}, Heinz and {Asorey}, Jacobo and {Battisti}, Andrew J. and {Bilicki}, Maciej and {Botteon}, Andrea and {Brown}, Michael and {Br{\"u}ggen}, Marcus and {Cowley}, Michael and {Dage}, Kristen and {Hale}, Catherine Laura and {Hardcastle}, Martin and {Kothes}, Roland and {Lazarevi{\'c}}, Sanja and {Lin}, Yen-Ting and {Luken}, Kieran and {Moss}, Jeremy and {Prathap}, P.~K. Jahang and {ur Rahman}, Syed Faisal and {Reiprich}, Thomas and {Riseley}, Christopher and {Salvato}, Mara and {Seymour}, Nicholas and {Shabala}, Stanislav and {Smith}, Daniel and {Vaccari}, Mattia and {van Loon}, Jacco Th. and {Wong}, O. Ivy Ivy and {Zainal Alsaberi}, Rami and {Asher}, Albany and {Ball}, Brianna and {Barbosa}, Davi and {Biava}, Nadia and {Bradley}, Aaron and {Carvajal}, Rodrigo and {Crawford}, Evan J. and {Galvin}, Timothy James and {Huynh}, Minh and {Leahy}, Denis and {Matute}, Israel and {Moss}, Vanessa and {Pappalardo}, Ciro and {Smeaton}, Zachary and {Velovi{\'c}}, Velibor and {Zafar}, Tayyaba},
        title = "{The Evolutionary Map of the Universe: A new radio atlas for the southern hemisphere sky}",
      journal = {\pasa},
         year = 2025,
        month = may,
       volume = {42},
          eid = {e071},
        pages = {e071},
          doi = {10.1017/pasa.2025.10042},
archivePrefix = {arXiv},
       eprint = {2505.08271},
 primaryClass = {astro-ph.GA},
       adsurl = {https://ui.adsabs.harvard.edu/abs/2025PASA...42...71H}
}

@ARTICLE{lee76,
       author = {{Lee}, L.~C. and {Jokipii}, J.~R.},
        title = "{The irregularity spectrum in interstellar space.}",
      journal = {\apj},
         year = 1976,
        month = jun,
       volume = {206},
        pages = {735-743},
          doi = {10.1086/154434},
       adsurl = {https://ui.adsabs.harvard.edu/abs/1976ApJ...206..735L}
}

@ARTICLE{smirnov_24,
       author = {{Zhu}, Xingzhi and {Zhang}, Zhehao and {Zhao}, Chengshi and {Li}, Bian and {Tong}, Minglei and {Gao}, Yuping and {Yang}, Tinggao},
        title = "{Research on establishing a joint time-scale of pulsar time and atomic time based on a wavelet analysis method}",
      journal = {\mnras},
         year = 2024,
        month = apr,
       volume = {529},
       number = {2},
        pages = {1082-1090},
          doi = {10.1093/mnras/stae331},
archivePrefix = {arXiv},
       eprint = {2312.12165},
 primaryClass = {astro-ph.IM},
       adsurl = {https://ui.adsabs.harvard.edu/abs/2024MNRAS.529.1082Z}
}

@ARTICLE{wang_24,
       author = {{Wang}, Ziteng and {Kaplan}, David L. and {Sengar}, Rahul and {Lenc}, Emil and {Zic}, Andrew and {Anumarlapudi}, Akash and {Gaensler}, B.~M. and {Hurley-Walker}, Natasha and {Murphy}, Tara and {Wang}, Yuanming},
        title = "{Discovery of a Young, Highly Scattered Pulsar PSR J1032-5804 with the Australian Square Kilometre Array Pathfinder}",
      journal = {\apj},
         year = 2024,
        month = feb,
       volume = {961},
       number = {2},
          eid = {175},
        pages = {175},
          doi = {10.3847/1538-4357/ad0fe8},
archivePrefix = {arXiv},
       eprint = {2311.14880},
 primaryClass = {astro-ph.HE},
       adsurl = {https://ui.adsabs.harvard.edu/abs/2024ApJ...961..175W}
}

@ARTICLE{ahmad_25,
       author = {{Ahmad}, A. and {Dai}, S. and {Lazarevi{\'c}}, S. and {Filipovi{\'c}}, M.~D. and {Johnston}, S. and {Kerr}, M. and {Li}, D. and {Maitra}, C. and {Manchester}, R.~N.},
        title = "{PSR J1631-4722: the discovery of a young and energetic pulsar in the supernova remnant G336.7+0.5}",
      journal = {\mnras},
         year = 2025,
        month = mar,
       volume = {537},
       number = {3},
        pages = {2868-2875},
          doi = {10.1093/mnras/staf181},
archivePrefix = {arXiv},
       eprint = {2412.11345},
 primaryClass = {astro-ph.HE},
       adsurl = {https://ui.adsabs.harvard.edu/abs/2025MNRAS.537.2868A}
}

@ARTICLE{mcconnell_20,
       author = {{McConnell}, D. and {Hale}, C.~L. and {Lenc}, E. and {Banfield}, J.~K. and {Heald}, George and {Hotan}, A.~W. and {Leung}, James K. and {Moss}, Vanessa A. and {Murphy}, Tara and {O'Brien}, Andrew and {Pritchard}, Joshua and {Raja}, Wasim and {Sadler}, Elaine M. and {Stewart}, Adam and {Thomson}, Alec J.~M. and {Whiting}, M. and {Allison}, James R. and {Amy}, S.~W. and {Anderson}, C. and {Ball}, Lewis and {Bannister}, Keith W. and {Bell}, Martin and {Bock}, Douglas C. -J. and {Bolton}, Russ and {Bunton}, J.~D. and {Chippendale}, A.~P. and {Collier}, J.~D. and {Cooray}, F.~R. and {Cornwell}, T.~J. and {Diamond}, P.~J. and {Edwards}, P.~G. and {Gupta}, N. and {Hayman}, Douglas B. and {Heywood}, Ian and {Jackson}, C.~A. and {Koribalski}, B{\"a}rbel S. and {Lee-Waddell}, Karen and {McClure-Griffiths}, N.~M. and {Ng}, Alan and {Norris}, Ray P. and {Phillips}, Chris and {Reynolds}, John E. and {Roxby}, Daniel N. and {Schinckel}, Antony E.~T. and {Shields}, Matt and {Tremblay}, Chenoa and {Tzioumis}, A. and {Voronkov}, M.~A. and {Westmeier}, Tobias},
        title = "{The Rapid ASKAP Continuum Survey I: Design and first results}",
      journal = {\pasa},
         year = 2020,
        month = nov,
       volume = {37},
          eid = {e048},
        pages = {e048},
          doi = {10.1017/pasa.2020.41},
archivePrefix = {arXiv},
       eprint = {2012.00747},
 primaryClass = {astro-ph.IM},
       adsurl = {https://ui.adsabs.harvard.edu/abs/2020PASA...37...48M}
}

@ARTICLE{zic_24,
       author = {{Zic}, Andrew and {Wang}, Ziteng and {Lenc}, Emil and {Kaplan}, David L. and {Murphy}, Tara and {Ridolfi}, A. and {Sengar}, Rahul and {Hurley-Walker}, Natasha and {Dobie}, Dougal and {Leung}, James K. and {Pritchard}, Joshua and {Wang}, Yuanming},
        title = "{Discovery of radio eclipses from 4FGL J1646.5-4406: a new candidate redback pulsar binary}",
      journal = {\mnras},
         year = 2024,
        month = mar,
       volume = {528},
       number = {4},
        pages = {5730-5741},
          doi = {10.1093/mnras/stae033},
archivePrefix = {arXiv},
       eprint = {2312.00261},
 primaryClass = {astro-ph.HE},
       adsurl = {https://ui.adsabs.harvard.edu/abs/2024MNRAS.528.5730Z}
}

@ARTICLE{2013A&A...558A..33A,
       author = {{Astropy Collaboration} and {Robitaille}, Thomas P. and
         {Tollerud}, Erik J. and {Greenfield}, Perry and {Droettboom}, Michael and
         {Bray}, Erik and {Aldcroft}, Tom and {Davis}, Matt and
         {Ginsburg}, Adam and {Price-Whelan}, Adrian M. and
         {Kerzendorf}, Wolfgang E. and {Conley}, Alexander and {Crighton}, Neil and
         {Barbary}, Kyle and {Muna}, Demitri and {Ferguson}, Henry and
         {Grollier}, Fr{\'e}d{\'e}ric and {Parikh}, Madhura M. and
         {Nair}, Prasanth H. and {Unther}, Hans M. and {Deil}, Christoph and
         {Woillez}, Julien and {Conseil}, Simon and {Kramer}, Roban and
         {Turner}, James E.~H. and {Singer}, Leo and {Fox}, Ryan and
         {Weaver}, Benjamin A. and {Zabalza}, Victor and {Edwards}, Zachary I. and
         {Azalee Bostroem}, K. and {Burke}, D.~J. and {Casey}, Andrew R. and
         {Crawford}, Steven M. and {Dencheva}, Nadia and {Ely}, Justin and
         {Jenness}, Tim and {Labrie}, Kathleen and {Lim}, Pey Lian and
         {Pierfederici}, Francesco and {Pontzen}, Andrew and {Ptak}, Andy and
         {Refsdal}, Brian and {Servillat}, Mathieu and {Streicher}, Ole},
        title = "{Astropy: A community Python package for astronomy}",
      journal = {\aap},
         year = "2013",
        month = "Oct",
       volume = {558},
          eid = {A33},
        pages = {A33},
          doi = {10.1051/0004-6361/201322068},
archivePrefix = {arXiv},
       eprint = {1307.6212},
 primaryClass = {astro-ph.IM},
       adsurl = {https://ui.adsabs.harvard.edu/abs/2013A&A...558A..33A}
}

@ARTICLE{2018AJ....156..123A,
       author = {{Astropy Collaboration} and {Price-Whelan}, A.~M. and {Sip{\H{o}}cz}, B.~M. and {G{\"u}nther}, H.~M. and {Lim}, P.~L. and {Crawford}, S.~M. and {Conseil}, S. and {Shupe}, D.~L. and {Craig}, M.~W. and {Dencheva}, N. and {Ginsburg}, A. and {VanderPlas}, J.~T. and {Bradley}, L.~D. and {P{\'e}rez-Su{\'a}rez}, D. and {de Val-Borro}, M. and {Aldcroft}, T.~L. and {Cruz}, K.~L. and {Robitaille}, T.~P. and {Tollerud}, E.~J. and {Ardelean}, C. and {Babej}, T. and {Bach}, Y.~P. and {Bachetti}, M. and {Bakanov}, A.~V. and {Bamford}, S.~P. and {Barentsen}, G. and {Barmby}, P. and {Baumbach}, A. and {Berry}, K.~L. and {Biscani}, F. and {Boquien}, M. and {Bostroem}, K.~A. and {Bouma}, L.~G. and {Brammer}, G.~B. and {Bray}, E.~M. and {Breytenbach}, H. and {Buddelmeijer}, H. and {Burke}, D.~J. and {Calderone}, G. and {Cano Rodr{\'\i}guez}, J.~L. and {Cara}, M. and {Cardoso}, J.~V.~M. and {Cheedella}, S. and {Copin}, Y. and {Corrales}, L. and {Crichton}, D. and {D'Avella}, D. and {Deil}, C. and {Depagne}, {\'E}. and {Dietrich}, J.~P. and {Donath}, A. and {Droettboom}, M. and {Earl}, N. and {Erben}, T. and {Fabbro}, S. and {Ferreira}, L.~A. and {Finethy}, T. and {Fox}, R.~T. and {Garrison}, L.~H. and {Gibbons}, S.~L.~J. and {Goldstein}, D.~A. and {Gommers}, R. and {Greco}, J.~P. and {Greenfield}, P. and {Groener}, A.~M. and {Grollier}, F. and {Hagen}, A. and {Hirst}, P. and {Homeier}, D. and {Horton}, A.~J. and {Hosseinzadeh}, G. and {Hu}, L. and {Hunkeler}, J.~S. and {Ivezi{\'c}}, {\v{Z}}. and {Jain}, A. and {Jenness}, T. and {Kanarek}, G. and {Kendrew}, S. and {Kern}, N.~S. and {Kerzendorf}, W.~E. and {Khvalko}, A. and {King}, J. and {Kirkby}, D. and {Kulkarni}, A.~M. and {Kumar}, A. and {Lee}, A. and {Lenz}, D. and {Littlefair}, S.~P. and {Ma}, Z. and {Macleod}, D.~M. and {Mastropietro}, M. and {McCully}, C. and {Montagnac}, S. and {Morris}, B.~M. and {Mueller}, M. and {Mumford}, S.~J. and {Muna}, D. and {Murphy}, N.~A. and {Nelson}, S. and {Nguyen}, G.~H. and {Ninan}, J.~P. and {N{\"o}the}, M. and {Ogaz}, S. and {Oh}, S. and {Parejko}, J.~K. and {Parley}, N. and {Pascual}, S. and {Patil}, R. and {Patil}, A.~A. and {Plunkett}, A.~L. and {Prochaska}, J.~X. and {Rastogi}, T. and {Reddy Janga}, V. and {Sabater}, J. and {Sakurikar}, P. and {Seifert}, M. and {Sherbert}, L.~E. and {Sherwood-Taylor}, H. and {Shih}, A.~Y. and {Sick}, J. and {Silbiger}, M.~T. and {Singanamalla}, S. and {Singer}, L.~P. and {Sladen}, P.~H. and {Sooley}, K.~A. and {Sornarajah}, S. and {Streicher}, O. and {Teuben}, P. and {Thomas}, S.~W. and {Tremblay}, G.~R. and {Turner}, J.~E.~H. and {Terr{\'o}n}, V. and {van Kerkwijk}, M.~H. and {de la Vega}, A. and {Watkins}, L.~L. and {Weaver}, B.~A. and {Whitmore}, J.~B. and {Woillez}, J. and {Zabalza}, V. and {Astropy Contributors}},
        title = "{The Astropy Project: Building an Open-science Project and Status of the v2.0 Core Package}",
      journal = {\aj},
         year = 2018,
        month = sep,
       volume = {156},
       number = {3},
          eid = {123},
        pages = {123},
          doi = {10.3847/1538-3881/aabc4f},
archivePrefix = {arXiv},
       eprint = {1801.02634},
 primaryClass = {astro-ph.IM},
       adsurl = {https://ui.adsabs.harvard.edu/abs/2018AJ....156..123A}
}

@MISC{2022ascl.soft10001K,
       author = {{Kaplan}, David L.},
        title = "{PSS: Pulsar Survey Scraper}",
 howpublished = {Astrophysics Source Code Library, record ascl:2210.001},
         year = 2022,
        month = oct,
          eid = {ascl:2210.001},
        pages = {ascl:2210.001},
archivePrefix = {ascl},
       eprint = {2210.001},
       adsurl = {https://ui.adsabs.harvard.edu/abs/2022ascl.soft10001K}
}

@ARTICLE{Bjornsson_90,
       author = {{Bjornsson}, C. -I.},
        title = "{Circular polarization in compact extragalactic radio sources}",
      journal = {\mnras},
         year = 1990,
        month = jan,
       volume = {242},
        pages = {158-166},
          doi = {10.1093/mnras/242.2.158},
       adsurl = {https://ui.adsabs.harvard.edu/abs/1990MNRAS.242..158B}
}

@ARTICLE{best_23,
       author = {{Best}, P.~N. and {Kondapally}, R. and {Williams}, W.~L. and {Cochrane}, R.~K. and {Duncan}, K.~J. and {Hale}, C.~L. and {Haskell}, P. and {Ma{\l}ek}, K. and {McCheyne}, I. and {Smith}, D.~J.~B. and {Wang}, L. and {Botteon}, A. and {Bonato}, M. and {Bondi}, M. and {Calistro Rivera}, G. and {Gao}, F. and {G{\"u}rkan}, G. and {Hardcastle}, M.~J. and {Jarvis}, M.~J. and {Mingo}, B. and {Miraghaei}, H. and {Morabito}, L.~K. and {Nisbet}, D. and {Prandoni}, I. and {R{\"o}ttgering}, H.~J.~A. and {Sabater}, J. and {Shimwell}, T. and {Tasse}, C. and {van Weeren}, R.},
        title = "{The LOFAR Two-metre Sky Survey: Deep Fields data release 1. V. Survey description, source classifications, and host galaxy properties}",
      journal = {\mnras},
         year = 2023,
        month = aug,
       volume = {523},
       number = {2},
        pages = {1729-1755},
          doi = {10.1093/mnras/stad1308},
archivePrefix = {arXiv},
       eprint = {2305.05782},
 primaryClass = {astro-ph.GA},
       adsurl = {https://ui.adsabs.harvard.edu/abs/2023MNRAS.523.1729B}
}

@ARTICLE{saikia_88,
       author = {{Saikia}, D.~J. and {Salter}, C.~J.},
        title = "{Polarization properties of extragalactic radio sources.}",
      journal = {\araa},
         year = 1988,
        month = jan,
       volume = {26},
        pages = {93-144},
          doi = {10.1146/annurev.aa.26.090188.000521},
       adsurl = {https://ui.adsabs.harvard.edu/abs/1988ARA&A..26...93S}
}

@ARTICLE{wang_23,
       author = {{Wang}, P.~F. and {Han}, J.~L. and {Xu}, J. and {Wang}, C. and {Yan}, Y. and {Jing}, W.~C. and {Su}, W.~Q. and {Zhou}, D.~J. and {Wang}, T.},
        title = "{FAST Pulsar Database. I. Polarization Profiles of 682 Pulsars}",
      journal = {Research in Astronomy and Astrophysics},
         year = 2023,
        month = oct,
       volume = {23},
       number = {10},
          eid = {104002},
        pages = {104002},
          doi = {10.1088/1674-4527/acea1f},
archivePrefix = {arXiv},
       eprint = {2307.10340},
 primaryClass = {astro-ph.HE},
       adsurl = {https://ui.adsabs.harvard.edu/abs/2023RAA....23j4002W}
}

@ARTICLE{oswald_23,
       author = {{Oswald}, L.~S. and {Johnston}, S. and {Karastergiou}, A. and {Dai}, S. and {Kerr}, M. and {Lower}, M.~E. and {Manchester}, R.~N. and {Shannon}, R.~M. and {Sobey}, C. and {Weltevrede}, P.},
        title = "{Pulsar polarization: a broad-band population view with the Parkes Ultra-Wideband receiver}",
      journal = {\mnras},
         year = 2023,
        month = apr,
       volume = {520},
       number = {4},
        pages = {4961-4980},
          doi = {10.1093/mnras/stad070},
archivePrefix = {arXiv},
       eprint = {2301.05628},
 primaryClass = {astro-ph.HE},
       adsurl = {https://ui.adsabs.harvard.edu/abs/2023MNRAS.520.4961O}
}

@ARTICLE{driessen_24,
       author = {{Driessen}, Laura Nicole and {Pritchard}, Joshua and {Murphy}, Tara and {Heald}, George and {Robrade}, Jan and {Das}, Barnali and {Duchesne}, Stefan William and {Kaplan}, David L. and {Lenc}, Emil and {Lynch}, Christene R. and {Mitchell-Bolton}, Jackson and {Pope}, Benjamin J.~S. and {Rose}, Kovi and {Stelzer}, Beate and {Wang}, Yuanming and {Zic}, Andrew},
        title = "{The Sydney Radio Star Catalogue: Properties of radio stars at megahertz to gigahertz frequencies}",
      journal = {\pasa},
         year = 2024,
        month = nov,
       volume = {41},
          eid = {e084},
        pages = {e084},
          doi = {10.1017/pasa.2024.72},
archivePrefix = {arXiv},
       eprint = {2404.07418},
 primaryClass = {astro-ph.SR},
       adsurl = {https://ui.adsabs.harvard.edu/abs/2024PASA...41...84D}
}

@ARTICLE{lenc_18,
       author = {{Lenc}, Emil and {Murphy}, Tara and {Lynch}, C.~R. and {Kaplan}, D.~L. and {Zhang}, S.~N.},
        title = "{An all-sky survey of circular polarization at 200 MHz}",
      journal = {\mnras},
         year = 2018,
        month = aug,
       volume = {478},
       number = {2},
        pages = {2835-2849},
          doi = {10.1093/mnras/sty1304},
archivePrefix = {arXiv},
       eprint = {1805.05482},
 primaryClass = {astro-ph.GA},
       adsurl = {https://ui.adsabs.harvard.edu/abs/2018MNRAS.478.2835L}
}

@ARTICLE{pritchard_21,
       author = {{Pritchard}, Joshua and {Murphy}, Tara and {Zic}, Andrew and {Lynch}, Christene and {Heald}, George and {Kaplan}, David L. and {Anderson}, Craig and {Banfield}, Julie and {Hale}, Catherine and {Hotan}, Aidan and {Lenc}, Emil and {Leung}, James K. and {McConnell}, David and {Moss}, Vanessa A. and {Raja}, Wasim and {Stewart}, Adam J. and {Whiting}, Matthew},
        title = "{A circular polarization survey for radio stars with the Australian SKA Pathfinder}",
      journal = {\mnras},
         year = 2021,
        month = apr,
       volume = {502},
       number = {4},
        pages = {5438-5454},
          doi = {10.1093/mnras/stab299},
archivePrefix = {arXiv},
       eprint = {2102.01801},
 primaryClass = {astro-ph.SR},
       adsurl = {https://ui.adsabs.harvard.edu/abs/2021MNRAS.502.5438P}
}

@ARTICLE{anumarlapudi_23,
       author = {{Anumarlapudi}, Akash and {Ehlke}, Anna and {Jones}, Megan L. and {Kaplan}, David L. and {Dobie}, Dougal and {Lenc}, Emil and {Leung}, James K. and {Murphy}, Tara and {Pritchard}, Joshua and {Stewart}, Adam J. and {Sengar}, Rahul and {Anderson}, Craig and {Banfield}, Julie and {Heald}, George and {Hotan}, Aidan W. and {McConnell}, David and {Moss}, Vanessa A. and {Raja}, Wasim and {Whiting}, Matthew T.},
        title = "{Characterizing Pulsars Detected in the Rapid ASKAP Continuum Survey}",
      journal = {\apj},
         year = 2023,
        month = oct,
       volume = {956},
       number = {1},
          eid = {28},
        pages = {28},
          doi = {10.3847/1538-4357/aceb5d},
archivePrefix = {arXiv},
       eprint = {2308.00100},
 primaryClass = {astro-ph.HE},
       adsurl = {https://ui.adsabs.harvard.edu/abs/2023ApJ...956...28A}
}

@ARTICLE{wang_22,
       author = {{Wang}, Yuanming and {Murphy}, Tara and {Kaplan}, David L. and {Klinner-Teo}, Teresa and {Ridolfi}, Alessandro and {Bailes}, Matthew and {Crawford}, Fronefield and {Dai}, Shi and {Dobie}, Dougal and {Gaensler}, B.~M. and {Graber}, Vanessa and {Heywood}, Ian and {Lenc}, Emil and {Lorimer}, Duncan R. and {McLaughlin}, Maura A. and {O'Brien}, Andrew and {Pintaldi}, Sergio and {Pritchard}, Joshua and {Rea}, Nanda and {Ridley}, Joshua P. and {Ronchi}, Michele and {Shannon}, Ryan M. and {Sivakoff}, Gregory R. and {Stewart}, Adam and {Wang}, Ziteng and {Zic}, Andrew},
        title = "{Discovery of PSR J0523-7125 as a Circularly Polarized Variable Radio Source in the Large Magellanic Cloud}",
      journal = {\apj},
         year = 2022,
        month = may,
       volume = {930},
       number = {1},
          eid = {38},
        pages = {38},
          doi = {10.3847/1538-4357/ac61dc},
archivePrefix = {arXiv},
       eprint = {2205.00622},
 primaryClass = {astro-ph.HE},
       adsurl = {https://ui.adsabs.harvard.edu/abs/2022ApJ...930...38W}
}

@ARTICLE{sobey_22,
       author = {{Sobey}, C. and {Bassa}, C.~G. and {O'Sullivan}, S.~P. and {Callingham}, J.~R. and {Tan}, C.~M. and {Hessels}, J.~W.~T. and {Kondratiev}, V.~I. and {Stappers}, B.~W. and {Tiburzi}, C. and {Heald}, G. and {Shimwell}, T. and {Breton}, R.~P. and {Kirwan}, M. and {Vedantham}, H.~K. and {Carretti}, E. and {Grie{\ss}meier}, J. -M. and {Haverkorn}, M. and {Karastergiou}, A.},
        title = "{Searching for pulsars associated with polarised point sources using LOFAR: Initial discoveries from the TULIPP project}",
      journal = {\aap},
         year = 2022,
        month = may,
       volume = {661},
          eid = {A87},
        pages = {A87},
          doi = {10.1051/0004-6361/202142636},
archivePrefix = {arXiv},
       eprint = {2203.08331},
 primaryClass = {astro-ph.HE},
       adsurl = {https://ui.adsabs.harvard.edu/abs/2022A&A...661A..87S}
}

@ARTICLE{jankowski_18,
       author = {{Jankowski}, F. and {van Straten}, W. and {Keane}, E.~F. and {Bailes}, M. and {Barr}, E.~D. and {Johnston}, S. and {Kerr}, M.},
        title = "{Spectral properties of 441 radio pulsars}",
      journal = {\mnras},
         year = 2018,
        month = feb,
       volume = {473},
       number = {4},
        pages = {4436-4458},
          doi = {10.1093/mnras/stx2476},
archivePrefix = {arXiv},
       eprint = {1709.08864},
 primaryClass = {astro-ph.HE},
       adsurl = {https://ui.adsabs.harvard.edu/abs/2018MNRAS.473.4436J}
}

@ARTICLE{thompson_94,
       author = {{Thompson}, C. and {Blandford}, R.~D. and {Evans}, Charles R. and {Phinney}, E.~S.},
        title = "{Physical Processes in Eclipsing Pulsars: Eclipse Mechanisms and Diagnostics}",
      journal = {\apj},
         year = 1994,
        month = feb,
       volume = {422},
        pages = {304},
          doi = {10.1086/173728},
       adsurl = {https://ui.adsabs.harvard.edu/abs/1994ApJ...422..304T}
}

@ARTICLE{rickett_77,
       author = {{Rickett}, B.~J.},
        title = "{Interstellar scattering and scintillation of radio waves.}",
      journal = {\araa},
         year = 1977,
        month = jan,
       volume = {15},
        pages = {479-504},
          doi = {10.1146/annurev.aa.15.090177.002403},
       adsurl = {https://ui.adsabs.harvard.edu/abs/1977ARA&A..15..479R}
}

@ARTICLE{2021PASA...38...38P,
       author = {{Price}, D.~C. and {Flynn}, C. and {Deller}, A.},
        title = "{A comparison of Galactic electron density models using PyGEDM}",
      journal = {\pasa},
         year = 2021,
        month = aug,
       volume = {38},
          eid = {e038},
        pages = {e038},
          doi = {10.1017/pasa.2021.33},
archivePrefix = {arXiv},
       eprint = {2106.15816},
 primaryClass = {astro-ph.GA},
       adsurl = {https://ui.adsabs.harvard.edu/abs/2021PASA...38...38P}
}

@ARTICLE{2004PASA...21..302H,
       author = {{Hotan}, A.~W. and {van Straten}, W. and {Manchester}, R.~N.},
        title = "{PSRCHIVE and PSRFITS: An Open Approach to Radio Pulsar Data Storage and Analysis}",
      journal = {\pasa},
         year = 2004,
        month = jan,
       volume = {21},
       number = {3},
        pages = {302-309},
          doi = {10.1071/AS04022},
archivePrefix = {arXiv},
       eprint = {astro-ph/0404549},
 primaryClass = {astro-ph},
       adsurl = {https://ui.adsabs.harvard.edu/abs/2004PASA...21..302H}
}

@ARTICLE{Lazarevi_24,
       author = {{Lazarevi{\'c}}, Sanja and {Filipovi{\'c}}, Miroslav D. and {Dai}, Shi and {Kothes}, Roland and {Ahmad}, Adeel and {Alsaberi}, Rami Z.~E. and {Balzan}, Joel C.~F. and {Barnes}, Luke A. and {Cotton}, William D. and {Edwards}, Philip G. and {Gordon}, Yjan A. and {Haberl}, Frank and {Hopkins}, Andrew M. and {Koribalski}, B{\"a}rbel S. and {Leahy}, Denis and {Maitra}, Chandreyee and {Mi{\'c}i{\'c}}, Marko and {Rowell}, Gavin and {Sasaki}, Manami and {Tothill}, Nicholas F.~H. and {Umana}, Grazia and {Velovi{\'c}}, Velibor},
        title = "{Fast as Potoroo: Radio continuum detection of a bow-shock pulsar wind nebula powered by pulsar J1638-4713}",
      journal = {\pasa},
         year = 2024,
        month = may,
       volume = {41},
          eid = {e032},
        pages = {e032},
          doi = {10.1017/pasa.2024.13},
archivePrefix = {arXiv},
       eprint = {2312.06961},
 primaryClass = {astro-ph.HE},
       adsurl = {https://ui.adsabs.harvard.edu/abs/2024PASA...41...32L}
}

@ARTICLE{sengar_htru_25,
  author       = {Sengar, R. and Bailes, M. and Balakrishnan, V. and Barr, E.~D. and
                  Bhat, N.~D.~R. and Burgay, M. and Bernadich, M.~C.~I. and
                  Cameron, A.~D. and Champion, D.~J. and Chen, W. and
                  Flynn, C.~M.~L. and Jameson, A. and Johnston, S. and
                  Keith, M.~J. and Kramer, M. and Morello, V. and Ng, C. and
                  Possenti, A. and Stevenson, S. and Shannon, R.~M. and
                  van Straten, W. and Wongphechauxsorn, J.},
  title        = {The High Time Resolution Universe Pulsar Survey - XIX. A coherent GPU-accelerated reprocessing and the discovery of 71 pulsars in the Southern Galactic plane},
  journal      = {\mnras},
  year         = {2025},
  month        = feb,
  volume       = {536},
  number       = {4},
  pages        = {3159--3176},
  doi          = {10.1093/mnras/stae2716},
  archivePrefix= {arXiv},
  eprint       = {2412.07104},
  primaryClass = {astro-ph.HE},
  adsurl       = {https://ui.adsabs.harvard.edu/abs/2025MNRAS.536.3159S}
}

@ARTICLE{sengar_23,
       author = {{Sengar}, R. and {Bailes}, M. and {Balakrishnan}, V. and {Bernadich}, M.~C. i. and {Burgay}, M. and {Barr}, E.~D. and {Flynn}, C.~M.~L. and {Stevenson}, R. Shannon S. and {Wongphechauxsorn}, J.},
        title = "{Discovery of 37 new pulsars through GPU-accelerated reprocessing of archival data of the Parkes multibeam pulsar survey}",
      journal = {\mnras},
         year = 2023,
        month = jun,
       volume = {522},
       number = {1},
        pages = {1071-1090},
          doi = {10.1093/mnras/stad508},
archivePrefix = {arXiv},
       eprint = {2302.00255},
 primaryClass = {astro-ph.HE},
       adsurl = {https://ui.adsabs.harvard.edu/abs/2023MNRAS.522.1071S}
}

@ARTICLE{jing_2025,
       author = {{Jing}, W.~C. and {Han}, J.~L. and {Wang}, C. and {Wang}, P.~F. and {Wang}, T. and {Cai}, N.~N. and {Xu}, J. and {Yang}, Z.~L. and {Zhou}, D.~J. and {Yan}, Yi and {Su}, W.~Q. and {Gao}, X.~Y. and {Xie}, L.},
        title = "{FAST Pulsar Database: II. Scattering profiles of 122 Pulsars}",
      journal = {arXiv e-prints},
         year = 2025,
        month = jun,
          eid = {arXiv:2506.14519},
        pages = {arXiv:2506.14519},
          doi = {10.48550/arXiv.2506.14519},
archivePrefix = {arXiv},
       eprint = {2506.14519},
 primaryClass = {astro-ph.HE},
       adsurl = {https://ui.adsabs.harvard.edu/abs/2025arXiv250614519J}
}

@ARTICLE{camilo_21,
       author = {{Camilo}, F. and {Ransom}, S.~M. and {Halpern}, J.~P. and {Roshi}, D. Anish},
        title = "{Radio Detection of PSR J1813-1749 in HESS J1813-178: The Most Scattered Pulsar Known}",
      journal = {\apj},
         year = 2021,
        month = aug,
       volume = {917},
       number = {2},
          eid = {67},
        pages = {67},
          doi = {10.3847/1538-4357/ac0720},
archivePrefix = {arXiv},
       eprint = {2106.00386},
 primaryClass = {astro-ph.HE},
       adsurl = {https://ui.adsabs.harvard.edu/abs/2021ApJ...917...67C}
}

@ARTICLE{camilo_11,
       author = {{Camilo}, F. and {Ransom}, S.~M. and {Chatterjee}, S. and {Johnston}, S. and {Demorest}, P.},
        title = "{PSR J1841-0500: A Radio Pulsar That Mostly is Not There}",
      journal = {\apj},
         year = 2012,
        month = feb,
       volume = {746},
       number = {1},
          eid = {63},
        pages = {63},
          doi = {10.1088/0004-637X/746/1/63},
archivePrefix = {arXiv},
       eprint = {1111.5870},
 primaryClass = {astro-ph.GA},
       adsurl = {https://ui.adsabs.harvard.edu/abs/2012ApJ...746...63C}
}

@ARTICLE{cordes_22,
       author = {{Cordes}, J.~M. and {Ocker}, Stella Koch and {Chatterjee}, Shami},
        title = "{Redshift Estimation and Constraints on Intergalactic and Interstellar Media from Dispersion and Scattering of Fast Radio Bursts}",
      journal = {\apj},
         year = 2022,
        month = jun,
       volume = {931},
       number = {2},
          eid = {88},
        pages = {88},
          doi = {10.3847/1538-4357/ac6873},
archivePrefix = {arXiv},
       eprint = {2108.01172},
 primaryClass = {astro-ph.HE},
       adsurl = {https://ui.adsabs.harvard.edu/abs/2022ApJ...931...88C}
}

@ARTICLE{ppta_13,
       author = {{Manchester}, R.~N. and {Hobbs}, G. and {Bailes}, M. and {Coles}, W.~A. and {van Straten}, W. and {Keith}, M.~J. and {Shannon}, R.~M. and {Bhat}, N.~D.~R. and {Brown}, A. and {Burke-Spolaor}, S.~G. and {Champion}, D.~J. and {Chaudhary}, A. and {Edwards}, R.~T. and {Hampson}, G. and {Hotan}, A.~W. and {Jameson}, A. and {Jenet}, F.~A. and {Kesteven}, M.~J. and {Khoo}, J. and {Kocz}, J. and {Maciesiak}, K. and {Oslowski}, S. and {Ravi}, V. and {Reynolds}, J.~R. and {Sarkissian}, J.~M. and {Verbiest}, J.~P.~W. and {Wen}, Z.~L. and {Wilson}, W.~E. and {Yardley}, D. and {Yan}, W.~M. and {You}, X.~P.},
        title = "{The Parkes Pulsar Timing Array Project}",
      journal = {\pasa},
         year = 2013,
        month = jan,
       volume = {30},
          eid = {e017},
        pages = {e017},
          doi = {10.1017/pasa.2012.017},
archivePrefix = {arXiv},
       eprint = {1210.6130},
 primaryClass = {astro-ph.IM},
       adsurl = {https://ui.adsabs.harvard.edu/abs/2013PASA...30...17M}
}

@INPROCEEDINGS{dai_18,
       author = {{Dai}, Shi and {Johnston}, Simon and {Hobbs}, George},
        title = "{Searching for pulsars in future radio continuum surveys}",
    booktitle = {Pulsar Astrophysics the Next Fifty Years},
         year = 2018,
       editor = {{Weltevrede}, P. and {Perera}, B.~B.~P. and {Preston}, L.~L. and {Sanidas}, S.},
       series = {IAU Symposium},
       volume = {337},
        month = aug,
        pages = {328-329},
          doi = {10.1017/S1743921317008833},
       adsurl = {https://ui.adsabs.harvard.edu/abs/2018IAUS..337..328D}
}

@ARTICLE{bhat_04,
       author = {{Bhat}, N.~D. Ramesh and {Cordes}, James M. and {Camilo}, Fernando and {Nice}, David J. and {Lorimer}, Duncan R.},
        title = "{Multifrequency Observations of Radio Pulse Broadening and Constraints on Interstellar Electron Density Microstructure}",
      journal = {\apj},
         year = 2004,
        month = apr,
       volume = {605},
       number = {2},
        pages = {759-783},
          doi = {10.1086/382680},
archivePrefix = {arXiv},
       eprint = {astro-ph/0401067},
 primaryClass = {astro-ph},
       adsurl = {https://ui.adsabs.harvard.edu/abs/2004ApJ...605..759B}
}

@ARTICLE{frail_24,
       author = {{Frail}, Dale A. and {Polisensky}, Emil and {Hyman}, Scott D. and {Cotton}, William D. and {Kassim}, Namir E. and {Silverstein}, Michele L. and {Sengar}, Rahul and {Kaplan}, David L. and {Calore}, Francesca and {Berteaud}, Joanna and {Clavel}, Ma{\"\i}ca and {Geyer}, Marisa and {Legodi}, Samuel and {Krishnan}, Vasaant and {Buchner}, Sarah and {Camilo}, Fernando},
        title = "{An Image-based Search for Pulsar Candidates in the MeerKAT Bulge Survey}",
      journal = {\apj},
         year = 2024,
        month = nov,
       volume = {975},
       number = {1},
          eid = {34},
        pages = {34},
          doi = {10.3847/1538-4357/ad74fd},
archivePrefix = {arXiv},
       eprint = {2407.01773},
 primaryClass = {astro-ph.HE},
       adsurl = {https://ui.adsabs.harvard.edu/abs/2024ApJ...975...34F}
}

@ARTICLE{backer_82,
       author = {{Backer}, D.~C. and {Kulkarni}, S.~R. and {Heiles}, C. and {Davis}, M.~M. and {Goss}, W.~M.},
        title = "{A millisecond pulsar}",
      journal = {\nat},
         year = 1982,
        month = dec,
       volume = {300},
       number = {5893},
        pages = {615-618},
          doi = {10.1038/300615a0},
       adsurl = {https://ui.adsabs.harvard.edu/abs/1982Natur.300..615B}
}

@ARTICLE{murphy_13,
       author = {{Murphy}, Tara and {Chatterjee}, Shami and {Kaplan}, David L. and {Banyer}, Jay and {Bell}, Martin E. and {Bignall}, Hayley E. and {Bower}, Geoffrey C. and {Cameron}, Robert A. and {Coward}, David M. and {Cordes}, James M. and {Croft}, Steve and {Curran}, James R. and {Djorgovski}, S.~G. and {Farrell}, Sean A. and {Frail}, Dale A. and {Gaensler}, B.~M. and {Galloway}, Duncan K. and {Gendre}, Bruce and {Green}, Anne J. and {Hancock}, Paul J. and {Johnston}, Simon and {Kamble}, Atish and {Law}, Casey J. and {Lazio}, T. Joseph W. and {Lo}, Kitty K. and {Macquart}, Jean-Pierre and {Rea}, Nanda and {Rebbapragada}, Umaa and {Reynolds}, Cormac and {Ryder}, Stuart D. and {Schmidt}, Brian and {Soria}, Roberto and {Stairs}, Ingrid H. and {Tingay}, Steven J. and {Torkelsson}, Ulf and {Wagstaff}, Kiri and {Walker}, Mark and {Wayth}, Randall B. and {Williams}, Peter K.~G.},
        title = "{VAST: An ASKAP Survey for Variables and Slow Transients}",
      journal = {\pasa},
         year = 2013,
        month = feb,
       volume = {30},
          eid = {e006},
        pages = {e006},
          doi = {10.1017/pasa.2012.006},
archivePrefix = {arXiv},
       eprint = {1207.1528},
 primaryClass = {astro-ph.IM},
       adsurl = {https://ui.adsabs.harvard.edu/abs/2013PASA...30....6M}
}

@ARTICLE{ne2001,
       author = {{Cordes}, J.~M. and {Lazio}, T.~J.~W.},
        title = "{NE2001.I. A New Model for the Galactic Distribution of Free Electrons and its Fluctuations}",
      journal = {arXiv e-prints},
         year = 2002,
        month = jul,
          eid = {astro-ph/0207156},
        pages = {astro-ph/0207156},
          doi = {10.48550/arXiv.astro-ph/0207156},
archivePrefix = {arXiv},
       eprint = {astro-ph/0207156},
 primaryClass = {astro-ph},
       adsurl = {https://ui.adsabs.harvard.edu/abs/2002astro.ph..7156C}
}

@ARTICLE{pint_21,
       author = {{Luo}, Jing and {Ransom}, Scott and {Demorest}, Paul and {Ray}, Paul S. and {Archibald}, Anne and {Kerr}, Matthew and {Jennings}, Ross J. and {Bachetti}, Matteo and {van Haasteren}, Rutger and {Champagne}, Chloe A. and {Colen}, Jonathan and {Phillips}, Camryn and {Zimmerman}, Josef and {Stovall}, Kevin and {Lam}, Michael T. and {Jenet}, Fredrick A.},
        title = "{PINT: A Modern Software Package for Pulsar Timing}",
      journal = {\apj},
         year = 2021,
        month = apr,
       volume = {911},
       number = {1},
          eid = {45},
        pages = {45},
          doi = {10.3847/1538-4357/abe62f},
archivePrefix = {arXiv},
       eprint = {2012.00074},
 primaryClass = {astro-ph.IM},
       adsurl = {https://ui.adsabs.harvard.edu/abs/2021ApJ...911...45L}
}

@ARTICLE{psrchive_12,
       author = {{van Straten}, Willem and {Demorest}, Paul and {Oslowski}, Stefan},
        title = "{Pulsar Data Analysis with PSRCHIVE}",
      journal = {Astronomical Research and Technology},
         year = 2012,
        month = jul,
       volume = {9},
       number = {3},
        pages = {237-256},
          doi = {10.48550/arXiv.1205.6276},
archivePrefix = {arXiv},
       eprint = {1205.6276},
 primaryClass = {astro-ph.IM},
       adsurl = {https://ui.adsabs.harvard.edu/abs/2012AR&T....9..237V}
}

@ARTICLE{hobbs_20,
       author = {{Hobbs}, George and {Manchester}, Richard N. and {Dunning}, Alex and {Jameson}, Andrew and {Roberts}, Paul and {George}, Daniel and {Green}, J.~A. and {Tuthill}, John and {Toomey}, Lawrence and {Kaczmarek}, Jane F. and {Mader}, Stacy and {Marquarding}, Malte and {Ahmed}, Azeem and {Amy}, Shaun W. and {Bailes}, Matthew and {Beresford}, Ron and {Bhat}, N.~D.~R. and {Bock}, Douglas C. -J. and {Bourne}, Michael and {Bowen}, Mark and {Brothers}, Michael and {Cameron}, Andrew D. and {Carretti}, Ettore and {Carter}, Nick and {Castillo}, Santy and {Chekkala}, Raji and {Cheng}, Wan and {Chung}, Yoon and {Craig}, Daniel A. and {Dai}, Shi and {Dawson}, Joanne and {Dempsey}, James and {Doherty}, Paul and {Dong}, Bin and {Edwards}, Philip and {Ergesh}, Tuohutinuer and {Gao}, Xuyang and {Han}, JinLin and {Hayman}, Douglas and {Indermuehle}, Balthasar and {Jeganathan}, Kanapathippillai and {Johnston}, Simon and {Kanoniuk}, Henry and {Kesteven}, Michael and {Kramer}, Michael and {Leach}, Mark and {Mcintyre}, Vince and {Moss}, Vanessa and {Os{\l}owski}, Stefan and {Phillips}, Chris and {Pope}, Nathan and {Preisig}, Brett and {Price}, Daniel and {Reeves}, Ken and {Reilly}, Les and {Reynolds}, John and {Robishaw}, Tim and {Roush}, Peter and {Ruckley}, Tim and {Sadler}, Elaine and {Sarkissian}, John and {Severs}, Sean and {Shannon}, Ryan and {Smart}, Ken and {Smith}, Malcolm and {Smith}, Stephanie and {Sobey}, Charlotte and {Staveley-Smith}, Lister and {Tzioumis}, Anastasios and {van Straten}, Willem and {Wang}, Nina and {Wen}, Linqing and {Whiting}, Matthew},
        title = "{An ultra-wide bandwidth (704 to 4 032 MHz) receiver for the Parkes radio telescope}",
      journal = {\pasa},
         year = 2020,
        month = apr,
       volume = {37},
          eid = {e012},
        pages = {e012},
          doi = {10.1017/pasa.2020.2},
archivePrefix = {arXiv},
       eprint = {1911.00656},
 primaryClass = {astro-ph.IM},
       adsurl = {https://ui.adsabs.harvard.edu/abs/2020PASA...37...12H}
}

@ARTICLE{ymw16,
       author = {{Yao}, J.~M. and {Manchester}, R.~N. and {Wang}, N.},
        title = "{A New Electron-density Model for Estimation of Pulsar and FRB Distances}",
      journal = {\apj},
         year = 2017,
        month = jan,
       volume = {835},
       number = {1},
          eid = {29},
        pages = {29},
          doi = {10.3847/1538-4357/835/1/29},
archivePrefix = {arXiv},
       eprint = {1610.09448},
 primaryClass = {astro-ph.GA},
       adsurl = {https://ui.adsabs.harvard.edu/abs/2017ApJ...835...29Y}
}

@ARTICLE{you_06,
       author = {{You}, Xiao-Peng and {Han}, Jin-lin},
        title = "{Circular Polarization in Pulsar Integrated Profiles: Updates}",
      journal = {\cjaa},
         year = 2006,
        month = apr,
       volume = {6},
       number = {2},
        pages = {237-246},
          doi = {10.1088/1009-9271/6/2/11},
archivePrefix = {arXiv},
       eprint = {astro-ph/0603597},
 primaryClass = {astro-ph},
       adsurl = {https://ui.adsabs.harvard.edu/abs/2006ChJAA...6..237Y}
}

@ARTICLE{murphy_21,
       author = {{Murphy}, Tara and {Kaplan}, David L. and {Stewart}, Adam J. and {O'Brien}, Andrew and {Lenc}, Emil and {Pintaldi}, Sergio and {Pritchard}, Joshua and {Dobie}, Dougal and {Fox}, Archibald and {Leung}, James K. and {An}, Tao and {Bell}, Martin E. and {Broderick}, Jess W. and {Chatterjee}, Shami and {Dai}, Shi and {d'Antonio}, Daniele and {Doyle}, Gerry and {Gaensler}, B.~M. and {Heald}, George and {Horesh}, Assaf and {Jones}, Megan L. and {McConnell}, David and {Moss}, Vanessa A. and {Raja}, Wasim and {Ramsay}, Gavin and {Ryder}, Stuart and {Sadler}, Elaine M. and {Sivakoff}, Gregory R. and {Wang}, Yuanming and {Wang}, Ziteng and {Wheatland}, Michael S. and {Whiting}, Matthew and {Allison}, James R. and {Anderson}, C.~S. and {Ball}, Lewis and {Bannister}, K. and {Bock}, D.~C. -J. and {Bolton}, R. and {Bunton}, J.~D. and {Chekkala}, R. and {Chippendale}, A.~P. and {Cooray}, F.~R. and {Gupta}, N. and {Hayman}, D.~B. and {Jeganathan}, K. and {Koribalski}, B. and {Lee-Waddell}, K. and {Mahony}, Elizabeth K. and {Marvil}, J. and {McClure-Griffiths}, N.~M. and {Mirtschin}, P. and {Ng}, A. and {Pearce}, S. and {Phillips}, C. and {Voronkov}, M.~A.},
        title = "{The ASKAP Variables and Slow Transients (VAST) Pilot Survey}",
      journal = {\pasa},
         year = 2021,
        month = oct,
       volume = {38},
          eid = {e054},
        pages = {e054},
          doi = {10.1017/pasa.2021.44},
archivePrefix = {arXiv},
       eprint = {2108.06039},
 primaryClass = {astro-ph.HE},
       adsurl = {https://ui.adsabs.harvard.edu/abs/2021PASA...38...54M}
}

@ARTICLE{norris_emu_2011,
       author = {{Norris}, Ray P. and {Hopkins}, A.~M. and {Afonso}, J. and {Brown}, S. and {Condon}, J.~J. and {Dunne}, L. and {Feain}, I. and {Hollow}, R. and {Jarvis}, M. and {Johnston-Hollitt}, M. and {Lenc}, E. and {Middelberg}, E. and {Padovani}, P. and {Prandoni}, I. and {Rudnick}, L. and {Seymour}, N. and {Umana}, G. and {Andernach}, H. and {Alexander}, D.~M. and {Appleton}, P.~N. and {Bacon}, D. and {Banfield}, J. and {Becker}, W. and {Brown}, M.~J.~I. and {Ciliegi}, P. and {Jackson}, C. and {Eales}, S. and {Edge}, A.~C. and {Gaensler}, B.~M. and {Giovannini}, G. and {Hales}, C.~A. and {Hancock}, P. and {Huynh}, M.~T. and {Ibar}, E. and {Ivison}, R.~J. and {Kennicutt}, R. and {Kimball}, Amy E. and {Koekemoer}, A.~M. and {Koribalski}, B.~S. and {L{\'o}pez-S{\'a}nchez}, {\'A}. R. and {Mao}, M.~Y. and {Murphy}, T. and {Messias}, H. and {Pimbblet}, K.~A. and {Raccanelli}, A. and {Randall}, K.~E. and {Reiprich}, T.~H. and {Roseboom}, I.~G. and {R{\"o}ttgering}, H. and {Saikia}, D.~J. and {Sharp}, R.~G. and {Slee}, O.~B. and {Smail}, Ian and {Thompson}, M.~A. and {Urquhart}, J.~S. and {Wall}, J.~V. and {Zhao}, G. -B.},
        title = "{EMU: Evolutionary Map of the Universe}",
      journal = {\pasa},
         year = 2011,
        month = aug,
       volume = {28},
       number = {3},
        pages = {215-248},
          doi = {10.1071/AS11021},
archivePrefix = {arXiv},
       eprint = {1106.3219},
 primaryClass = {astro-ph.CO},
       adsurl = {https://ui.adsabs.harvard.edu/abs/2011PASA...28..215N}
}

@ARTICLE{vlass_20,
       author = {{Lacy}, M. and {Baum}, S.~A. and {Chandler}, C.~J. and {Chatterjee}, S. and {Clarke}, T.~E. and {Deustua}, S. and {English}, J. and {Farnes}, J. and {Gaensler}, B.~M. and {Gugliucci}, N. and {Hallinan}, G. and {Kent}, B.~R. and {Kimball}, A. and {Law}, C.~J. and {Lazio}, T.~J.~W. and {Marvil}, J. and {Mao}, S.~A. and {Medlin}, D. and {Mooley}, K. and {Murphy}, E.~J. and {Myers}, S. and {Osten}, R. and {Richards}, G.~T. and {Rosolowsky}, E. and {Rudnick}, L. and {Schinzel}, F. and {Sivakoff}, G.~R. and {Sjouwerman}, L.~O. and {Taylor}, R. and {White}, R.~L. and {Wrobel}, J. and {Andernach}, H. and {Beasley}, A.~J. and {Berger}, E. and {Bhatnager}, S. and {Birkinshaw}, M. and {Bower}, G.~C. and {Brandt}, W.~N. and {Brown}, S. and {Burke-Spolaor}, S. and {Butler}, B.~J. and {Comerford}, J. and {Demorest}, P.~B. and {Fu}, H. and {Giacintucci}, S. and {Golap}, K. and {G{\"u}th}, T. and {Hales}, C.~A. and {Hiriart}, R. and {Hodge}, J. and {Horesh}, A. and {Ivezi{\'c}}, {\v{Z}}. and {Jarvis}, M.~J. and {Kamble}, A. and {Kassim}, N. and {Liu}, X. and {Loinard}, L. and {Lyons}, D.~K. and {Masters}, J. and {Mezcua}, M. and {Moellenbrock}, G.~A. and {Mroczkowski}, T. and {Nyland}, K. and {O'Dea}, C.~P. and {O'Sullivan}, S.~P. and {Peters}, W.~M. and {Radford}, K. and {Rao}, U. and {Robnett}, J. and {Salcido}, J. and {Shen}, Y. and {Sobotka}, A. and {Witz}, S. and {Vaccari}, M. and {van Weeren}, R.~J. and {Vargas}, A. and {Williams}, P.~K.~G. and {Yoon}, I.},
        title = "{The Karl G. Jansky Very Large Array Sky Survey (VLASS). Science Case and Survey Design}",
      journal = {\pasp},
         year = 2020,
        month = mar,
       volume = {132},
       number = {1009},
          eid = {035001},
        pages = {035001},
          doi = {10.1088/1538-3873/ab63eb},
archivePrefix = {arXiv},
       eprint = {1907.01981},
 primaryClass = {astro-ph.IM},
       adsurl = {https://ui.adsabs.harvard.edu/abs/2020PASP..132c5001L}
}

@INPROCEEDINGS{fender_16,
       author = {{Fender}, R. and {Woudt}, P.~A. and {Corbel}, S. and {Coriat}, M. and {Daigne}, F. and {Falcke}, H. and {Girard}, J. and {Heywood}, I. and {Horesh}, A. and {Horrell}, J. and {Jonker}, P.~G. and {Joseph}, T. and {Kamble}, A. and {Knigge}, C. and {K{\"o}rding}, E. and {Kotze}, M. and {Kouveliotou}, C. and {Lynch}, C. and {Maccarone}, T. and {Meintjes}, P. and {Migliari}, S. and {Murphy}, T. and {Nagayama}, T. and {Nelemans}, G. and {Nicholson}, G. and {O'Brien}, T. and {Oodendaal}, A. and {Oozeer}, N. and {Osborne}, J. and {P{\'e}rez-Torres}, M. and {Ratcliffe}, S. and {Ribeiro}, V.~A.~R.~M. and {Rol}, E. and {Rushton}, A. and {Scaife}, A. and {Schurch}, M. and {Sivakoff}, G. and {Staley}, T. and {Steeghs}, D. and {Stewart}, I. and {Swinbank}, J.~D. and {Vergani}, S. and {Warner}, B. and {Wiersema}, K. and {Armstrong}, R. and {Groot}, P. and {McBride}, V. and {Miller-Jones}, J.~C.~A. and {Mooley}, K. and {Stappers}, B. and {Wijers}, R.~A.~M.~J. and {Bietenholz}, M. and {Blyth}, S. and {B{\"o}ttcher}, M. and {Buckley}, D. and {Charles}, P. and {Chomiuk}, L. and {Coppejans}, D. and {de Blok}, W.~J.~G. and {van der Heyden}, K. and {van der Horst}, A. and {van Soelen}, B.},
        title = "{ThunderKAT: The MeerKAT Large Survey Project for Image-Plane Radio Transients}",
    booktitle = {MeerKAT Science: On the Pathway to the SKA},
         year = 2016,
        month = jan,
          eid = {13},
        pages = {13},
          doi = {10.22323/1.277.0013},
archivePrefix = {arXiv},
       eprint = {1711.04132},
 primaryClass = {astro-ph.HE},
       adsurl = {https://ui.adsabs.harvard.edu/abs/2016mks..confE..13F}
}

@ARTICLE{padmanabh_23,
       author = {{Padmanabh}, P.~V. and {Barr}, E.~D. and {Sridhar}, S.~S. and {Rugel}, M.~R. and {Damas-Segovia}, A. and {Jacob}, A.~M. and {Balakrishnan}, V. and {Berezina}, M. and {Bernadich}, M.~C. and {Brunthaler}, A. and {Champion}, D.~J. and {Freire}, P.~C.~C. and {Khan}, S. and {Kl{\"o}ckner}, H. -R. and {Kramer}, M. and {Ma}, Y.~K. and {Mao}, S.~A. and {Men}, Y.~P. and {Menten}, K.~M. and {Sengupta}, S. and {Venkatraman Krishnan}, V. and {Wucknitz}, O. and {Wyrowski}, F. and {Bezuidenhout}, M.~C. and {Buchner}, S. and {Burgay}, M. and {Chen}, W. and {Clark}, C.~J. and {K{\"u}nkel}, L. and {Nieder}, L. and {Stappers}, B. and {Legodi}, L.~S. and {Nyamai}, M.~M.},
        title = "{The MPIfR-MeerKAT Galactic Plane Survey - I. System set-up and early results}",
      journal = {\mnras},
         year = 2023,
        month = sep,
       volume = {524},
       number = {1},
        pages = {1291-1315},
          doi = {10.1093/mnras/stad1900},
archivePrefix = {arXiv},
       eprint = {2303.09231},
 primaryClass = {astro-ph.HE},
       adsurl = {https://ui.adsabs.harvard.edu/abs/2023MNRAS.524.1291P}
}

@ARTICLE{manchester_05,
       author = {{Manchester}, R.~N. and {Hobbs}, G.~B. and {Teoh}, A. and {Hobbs}, M.},
        title = "{The Australia Telescope National Facility Pulsar Catalogue}",
      journal = {\aj},
         year = 2005,
        month = apr,
       volume = {129},
       number = {4},
        pages = {1993-2006},
          doi = {10.1086/428488},
archivePrefix = {arXiv},
       eprint = {astro-ph/0412641},
 primaryClass = {astro-ph},
       adsurl = {https://ui.adsabs.harvard.edu/abs/2005AJ....129.1993M}
}

@BOOK{handbook_04,
       author = {{Lorimer}, D.~R. and {Kramer}, M.},
        title = "{Handbook of Pulsar Astronomy}",
         year = 2004,
       volume = {4},
       adsurl = {https://ui.adsabs.harvard.edu/abs/2004hpa..book.....L}
}

@ARTICLE{staelin_69,
       author = {{Staelin}, D.~H.},
        title = "{Fast folding algorithm for detection of periodic pulse trains.}",
      journal = {IEEE Proceedings},
         year = 1969,
        month = jan,
       volume = {57},
        pages = {724-725},
          doi = {10.1109/PROC.1969.7051},
       adsurl = {https://ui.adsabs.harvard.edu/abs/1969IEEEP..57..724S}
}

@ARTICLE{sutton_ism_71,
       author = {{Sutton}, J.~M.},
        title = "{Scattering of pulsar radiation in the interstellar medium}",
      journal = {\mnras},
         year = 1971,
        month = jan,
       volume = {155},
        pages = {51},
          doi = {10.1093/mnras/155.1.51},
       adsurl = {https://ui.adsabs.harvard.edu/abs/1971MNRAS.155...51S}
}

@ARTICLE{keith_10,
       author = {{Keith}, M.~J. and {Jameson}, A. and {van Straten}, W. and {Bailes}, M. and {Johnston}, S. and {Kramer}, M. and {Possenti}, A. and {Bates}, S.~D. and {Bhat}, N.~D.~R. and {Burgay}, M. and {Burke-Spolaor}, S. and {D'Amico}, N. and {Levin}, L. and {McMahon}, Peter L. and {Milia}, S. and {Stappers}, B.~W.},
        title = "{The High Time Resolution Universe Pulsar Survey - I. System configuration and initial discoveries}",
      journal = {\mnras},
         year = 2010,
        month = dec,
       volume = {409},
       number = {2},
        pages = {619-627},
          doi = {10.1111/j.1365-2966.2010.17325.x},
archivePrefix = {arXiv},
       eprint = {1006.5744},
 primaryClass = {astro-ph.HE},
       adsurl = {https://ui.adsabs.harvard.edu/abs/2010MNRAS.409..619K}
}

@ARTICLE{manchester_01,
       author = {{Manchester}, R.~N. and {Lyne}, A.~G. and {Camilo}, F. and {Bell}, J.~F. and {Kaspi}, V.~M. and {D'Amico}, N. and {McKay}, N.~P.~F. and {Crawford}, F. and {Stairs}, I.~H. and {Possenti}, A. and {Kramer}, M. and {Sheppard}, D.~C.},
        title = "{The Parkes multi-beam pulsar survey - I. Observing and data analysis systems, discovery and timing of 100 pulsars}",
      journal = {\mnras},
         year = 2001,
        month = nov,
       volume = {328},
       number = {1},
        pages = {17-35},
          doi = {10.1046/j.1365-8711.2001.04751.x},
archivePrefix = {arXiv},
       eprint = {astro-ph/0106522},
 primaryClass = {astro-ph},
       adsurl = {https://ui.adsabs.harvard.edu/abs/2001MNRAS.328...17M}
}

@ARTICLE{cordes_06,
       author = {{Cordes}, J.~M. and {Freire}, P.~C.~C. and {Lorimer}, D.~R. and {Camilo}, F. and {Champion}, D.~J. and {Nice}, D.~J. and {Ramachandran}, R. and {Hessels}, J.~W.~T. and {Vlemmings}, W. and {van Leeuwen}, J. and {Ransom}, S.~M. and {Bhat}, N.~D.~R. and {Arzoumanian}, Z. and {McLaughlin}, M.~A. and {Kaspi}, V.~M. and {Kasian}, L. and {Deneva}, J.~S. and {Reid}, B. and {Chatterjee}, S. and {Han}, J.~L. and {Backer}, D.~C. and {Stairs}, I.~H. and {Deshpande}, A.~A. and {Faucher-Gigu{\`e}re}, C. -A.},
        title = "{Arecibo Pulsar Survey Using ALFA. I. Survey Strategy and First Discoveries}",
      journal = {\apj},
         year = 2006,
        month = jan,
       volume = {637},
       number = {1},
        pages = {446-455},
          doi = {10.1086/498335},
archivePrefix = {arXiv},
       eprint = {astro-ph/0509732},
 primaryClass = {astro-ph},
       adsurl = {https://ui.adsabs.harvard.edu/abs/2006ApJ...637..446C}
}

@ARTICLE{keane_18,
       author = {{Keane}, E.~F. and {Barr}, E.~D. and {Jameson}, A. and {Morello}, V. and {Caleb}, M. and {Bhandari}, S. and {Petroff}, E. and {Possenti}, A. and {Burgay}, M. and {Tiburzi}, C. and {Bailes}, M. and {Bhat}, N.~D.~R. and {Burke-Spolaor}, S. and {Eatough}, R.~P. and {Flynn}, C. and {Jankowski}, F. and {Johnston}, S. and {Kramer}, M. and {Levin}, L. and {Ng}, C. and {van Straten}, W. and {Krishnan}, V. Venkatraman},
        title = "{The SUrvey for Pulsars and Extragalactic Radio Bursts - I. Survey description and overview}",
      journal = {\mnras},
         year = 2018,
        month = jan,
       volume = {473},
       number = {1},
        pages = {116-135},
          doi = {10.1093/mnras/stx2126},
archivePrefix = {arXiv},
       eprint = {1706.04459},
 primaryClass = {astro-ph.IM},
       adsurl = {https://ui.adsabs.harvard.edu/abs/2018MNRAS.473..116K}
}

@ARTICLE{han_21,
       author = {{Han}, J.~L. and {Wang}, Chen and {Wang}, P.~F. and {Wang}, Tao and {Zhou}, D.~J. and {Sun}, Jing-Hai and {Yan}, Yi and {Su}, Wei-Qi and {Jing}, Wei-Cong and {Chen}, Xue and {Gao}, X.~Y. and {Hou}, Li-Gang and {Xu}, Jun and {Lee}, K.~J. and {Wang}, Na and {Jiang}, Peng and {Xu}, Ren-Xin and {Yan}, Jun and {Gan}, Heng-Qian and {Guan}, Xin and {Huang}, Wen-Jun and {Jiang}, Jin-Chen and {Li}, Hui and {Men}, Yun-Peng and {Sun}, Chun and {Wang}, Bo-Jun and {Wang}, H.~G. and {Wang}, Shuang-Qiang and {Xie}, Jin-Tao and {Xu}, Heng and {Yao}, Rui and {You}, Xiao-Peng and {Yu}, D.~J. and {Yuan}, Jian-Ping and {Yuen}, Rai and {Zhang}, Chun-Feng and {Zhu}, Yan},
        title = "{The FAST Galactic Plane Pulsar Snapshot survey: I. Project design and pulsar discoveries}",
      journal = {Research in Astronomy and Astrophysics},
         year = 2021,
        month = jun,
       volume = {21},
       number = {5},
          eid = {107},
        pages = {107},
          doi = {10.1088/1674-4527/21/5/107},
archivePrefix = {arXiv},
       eprint = {2105.08460},
 primaryClass = {astro-ph.HE},
       adsurl = {https://ui.adsabs.harvard.edu/abs/2021RAA....21..107H}
}

@ARTICLE{sanidas_19,
       author = {{Sanidas}, S. and {Cooper}, S. and {Bassa}, C.~G. and {Hessels}, J.~W.~T. and {Kondratiev}, V.~I. and {Michilli}, D. and {Stappers}, B.~W. and {Tan}, C.~M. and {van Leeuwen}, J. and {Cerrigone}, L. and {Fallows}, R.~A. and {Iacobelli}, M. and {Orr{\'u}}, E. and {Pizzo}, R.~F. and {Shulevski}, A. and {Toribio}, M.~C. and {ter Veen}, S. and {Zucca}, P. and {Bondonneau}, L. and {Grie{\ss}meier}, J. -M. and {Karastergiou}, A. and {Kramer}, M. and {Sobey}, C.},
        title = "{The LOFAR Tied-Array All-Sky Survey (LOTAAS): Survey overview and initial pulsar discoveries}",
      journal = {\aap},
         year = 2019,
        month = jun,
       volume = {626},
          eid = {A104},
        pages = {A104},
          doi = {10.1051/0004-6361/201935609},
archivePrefix = {arXiv},
       eprint = {1905.04977},
 primaryClass = {astro-ph.HE},
       adsurl = {https://ui.adsabs.harvard.edu/abs/2019A&A...626A.104S}
}

@ARTICLE{johnston_91,
       author = {{Johnston}, Helen M. and {Kulkarni}, Shrinivas R.},
        title = "{On the Detectability of Pulsars in Close Binary Systems}",
      journal = {\apj},
         year = 1991,
        month = feb,
       volume = {368},
        pages = {504},
          doi = {10.1086/169715},
       adsurl = {https://ui.adsabs.harvard.edu/abs/1991ApJ...368..504J}
}

@ARTICLE{maura_06,
       author = {{McLaughlin}, M.~A. and {Lyne}, A.~G. and {Lorimer}, D.~R. and {Kramer}, M. and {Faulkner}, A.~J. and {Manchester}, R.~N. and {Cordes}, J.~M. and {Camilo}, F. and {Possenti}, A. and {Stairs}, I.~H. and {Hobbs}, G. and {D'Amico}, N. and {Burgay}, M. and {O'Brien}, J.~T.},
        title = "{Transient radio bursts from rotating neutron stars}",
      journal = {\nat},
         year = 2006,
        month = feb,
       volume = {439},
       number = {7078},
        pages = {817-820},
          doi = {10.1038/nature04440},
archivePrefix = {arXiv},
       eprint = {astro-ph/0511587},
 primaryClass = {astro-ph},
       adsurl = {https://ui.adsabs.harvard.edu/abs/2006Natur.439..817M}
}
\bibliographystyle{aasjournal}

\end{document}